\documentclass[format=acmsmall, nonacm]{acmart}

\usepackage{booktabs} 
\usepackage{booktabs}
\usepackage{graphicx}
\usepackage{xcolor, subcaption}
\usepackage{url}
\usepackage{siunitx}
\usepackage{tabularx}
\usepackage{colortbl}
\usepackage{textcomp, color, soul}
\usepackage{color,soul}
\usepackage{amsmath, wrapfig}
\usepackage{tikz}
\usepackage{ctable}
\usepackage{tikzsymbols}
\usepackage{textcomp}
\usepackage{booktabs} 
\usepackage{courier}
\usepackage{booktabs}
\usepackage{graphicx}
\usepackage{xcolor}
\usepackage{multirow}

\usepackage{tcolorbox}

\usepackage[colorinlistoftodos]{todonotes}

\usepackage{fontawesome}    
\usepackage{fancyhdr}   
\fancypagestyle{firststyle}{%
  \fancyhf{}
  \fancyfoot[C]{\textcolor{gray}{\scriptsize{This manuscript is under review. Please write to vedantswain@gatech.edu for up-to-date information.}}}
}
\fancypagestyle{allstyle}{%
   \fancyfoot{}
  \fancyfoot[C]{\textcolor{gray}{\scriptsize{This manuscript is under review. Please write to vedantswain@gatech.edu for up-to-date information.}}}
}

\usepackage{url}
\usepackage{tabularx}
\usepackage{textcomp, color, soul, xcolor}
\usepackage{color,soul}
\usepackage{amsmath, wrapfig}
\usepackage{tikz}
\usepackage{ctable}
\usepackage{capt-of}
\usepackage{subcaption, balance}
\usepackage{balance,bm}
\usepackage [english]{babel}
\usepackage{url}
\usepackage [autostyle, english = american]{csquotes}
\definecolor{linkColor}{RGB}{6,125,233}
\definecolor{green}{rgb}{0.0, 0.65, 0.31}
\definecolor{bleudefrance}{rgb}{0.19, 0.55, 0.91}
\definecolor{ceruleanblue}{rgb}{0.16, 0.32, 0.75}
\definecolor{mediumblue}{rgb}{0.0, 0.0, 0.8}
\definecolor{grey}{HTML}{969696}
\definecolor{violet}{HTML}{756bb1}
\definecolor{dgrey}{HTML}{01665e}
\definecolor{lgrey}{HTML}{5ab4ac}
\definecolor{dgreen}{HTML}{005a32}
\definecolor{purple}{HTML}{8B008B}

\colorlet{editCol}{black}
\colorlet{editCol2}{black}

\definecolor{maskCol}{HTML}{c51b7d}
\definecolor{lrColor}{HTML}{8856a7}
\definecolor{trColor}{HTML}{d01c8b}
\definecolor{ctColor}{HTML}{4dac26}
\definecolor{brickred}{HTML}{f03b20}
\definecolor{improveCol}{HTML}{253494}
\definecolor{worsenCol}{HTML}{d7191c}
\definecolor{violet}{HTML}{6a51a3}

\definecolor{rubinered}{HTML}{CE0058}
\definecolor{midnightblue}{HTML}{191970}

\newcommand*{\textlabel}[2]{%
  \edef\@currentlabel{#1}
  \phantomsection
  #1\label{#2}
}

\colorlet{tableheadcolor}{gray!25} 
\colorlet{tablerowcolor}{gray!15} 
\colorlet{tablerowcolor2}{gray!12} 
\newcommand{\rowcollight}{\rowcolor{tablerowcolor2}} %

\newcommand{\edit}[1]{{\textcolor{editCol}{#1}}}
\newcommand{\revision}[1]{{\textcolor{editCol2}{#1}}}

\newcommand{\Mz}{$M_{0}$}
\newcommand{\Mpe}{$M_{PE}$}
\newcommand{\Mind}{$M_{iWF}$}
\newcommand{\Mgro}{$M_{gWF}$}

\newcommand{\Mindxgro}{$M_{iWF.gWF}$}

\renewcommand{\textrightarrow}{$\rightarrow$}

\colorlet{tableheadcolor}{gray!25} 
\colorlet{tablerowcolor}{gray!5} 

\definecolor{neutralCol}{HTML}{dd1c77}
\definecolor{neutralGreen}{HTML}{31a354}
\definecolor{NewBlue}{HTML}{1879ba}
\definecolor{bleudefrance}{rgb}{0.19, 0.55, 0.91}  
\definecolor{AfTrColor}{HTML}{0868ac}  
\definecolor{BfTrColor}{HTML}{a8ddb5}  

\definecolor{AfCtColor}{HTML}{b10026}  
\definecolor{BfCtColor}{HTML}{fd8d3c}

\newcommand{\iconcheck}{\textcolor{green}{\faCheckCircle}}
\newcommand{\iconnone}{\textcolor{grey}{\faMinusCircle}}

\usepackage[ruled]{algorithm2e} 

\makeatletter
\let\@authorsaddresses\@empty
\makeatother

\begin{document}
\title[Leveraging WiFi Network Logs to Infer Student Collocation]{Leveraging WiFi Network Logs to Infer Student Collocation and its Relationship with Academic Performance}


\author{Vedant Das Swain}
\email{vedantswain@gatech.edu}
\affiliation{%
  \institution{Georgia Institute of Technology}
  \country{USA}
}
\author{Hyeokhyen Kwon}
\email{hyeokhyen@gatech.edu}
\affiliation{%
  \institution{Georgia Institute of Technology}
  \country{USA}
}
\author{Sonia Sargolzaei}
\email{sonia.s@gatech.edu}
\affiliation{%
  \institution{Georgia Institute of Technology}
  \country{USA}
}
\author{Bahador Saket}
\email{saket@gatech.edu}
\affiliation{%
  \institution{Georgia Institute of Technology}
  \country{USA}
}
\author{Mehrab Bin Morshed}
\email{mehrab.morshed@gatech.edu}
\affiliation{%
  \institution{Georgia Institute of Technology}
  \country{USA}
}
\author{Kathy Tran}
\email{kathy.tran@gatech.edu}
\affiliation{%
  \institution{Georgia Institute of Technology}
  \country{USA}
}
\author{Devashru Patel}
\email{devashru@gatech.edu}
\affiliation{%
  \institution{Georgia Institute of Technology}
  \country{USA}
}
\author{Yexin Tian}
\email{yexintian.morshed@gatech.edu}
\affiliation{%
  \institution{Georgia Institute of Technology}
  \country{USA}
}
\author{Joshua Philipose}
\email{jphilipose3@gatech.edu}
\affiliation{%
  \institution{Georgia Institute of Technology}
  \country{USA}
}
\author{Yulai Cui}
\email{ycui96@gatech.edu}
\affiliation{%
  \institution{Georgia Institute of Technology}
  \country{USA}
}
\author{Thomas Pl\"{o}tz}
\email{thomas.ploetz@gatech.edu}
\affiliation{%
  \institution{Georgia Institute of Technology}
  \country{USA}
}
\author{Munmun De Choudhury}
\email{munmun.choudhury@cc.gatech.edu}
\affiliation{%
  \institution{Georgia Institute of Technology}
  \country{USA}
}
\author{Gregory D. Abowd}
\email{abowd@gatech.edu}
\affiliation{%
  \institution{Georgia Institute of Technology}
  \country{USA}
}

\renewcommand\shortauthors{Das Swain, V. et al}

\begin{abstract}
\edit{
A comprehensive understanding of collocation can help understand performance outcomes. For university cohorts, this needs data that describes large groups over a long period. 
Harnessing user devices to infer this, while tempting, is challenged by privacy concerns, power consumption, and maintenance issues. Alternatively, embedding new sensors in the environment is limited by the expense of covering the entire campus. 
We investigate the feasibility of leveraging WiFi association logs for this purpose. While these provide coarse approximations of location, these are easily obtainable and depict multiple users on campus over a semester.  
We explore how these coarse collocations are related to individual performance. Specifically, we inspect the association between individual performance and the collocation behaviors of project group members. We study 163 students (in 54 project groups) over 14 weeks. After describing how we determine collocation with the WiFi logs, we present a study to analyze how collocation within groups relates to a student's final score. We find collocation behaviors show a significant correlation (\textit{Pearson's r = 0.24})  with  performance --- better than both peer feedback or individual behaviors like attendance.  Finally, we discuss how repurposing WiFi logs can facilitate applications for domains like mental wellbeing and physical health.
}

\end{abstract}

%
%

%
%

\maketitle
\thispagestyle{firststyle} 
\pagestyle{allstyle}

\section{Introduction}


Humans are social by nature; their functioning is related to behaviors that are interlinked with those of others~\cite{homans1974social}.  One of the ways these behaviors manifest is when people in the same physical space take mutually-oriented actions~\cite{rummel1976understanding}. 
\revision{Especially in the context of work, it has been observed that being collocated in the same space provides common artifacts for reference and helps collaborators coordinate their effort~\cite{olson2000distance}. Additionally, collocation provides the opportunity for synchronous interactions through multiple channels --- voice, expressions, gestures and body posture --- and for impromptu interactions that strengthen social ties.}
\revision{While the benefits of collocation have been extensively documented in the context of information workers~\cite{olson2000distance,mark2005no,kozlowski2006enhancing,hinds2003out,geister2006effects},}
understanding how students ~\revision{collocate} can help campus stakeholders gain valuable insights to support academic outcomes. 
\edit{However, evaluating these behaviors with traditional surveys} \revision{is obtrusive and does not scale to represent dynamic human functioning.}

\edit{The passive sensing community has introduced many automated and unobtrusive sensing methods to capture social interactions~\cite{lukowicz2011context,olguin2008sensible,eagle2006reality,wang2015smartgpa,das2018groupsense} \revision{between collocated individuals}. 
However, approaches that require specialized client devices~\cite{eagle2006reality,olguin2008sensible,nguyen2013extraction}
have several limitations that constrain consistent data collection over significant periods of time. This includes technical challenges such as evolving manufacturer specifications that can critically disrupt data collection during mid-study with an update. In fact, the privacy concerns of installing such sensing firmware has also limited the meaningful data-streams available to users~\cite{shilton2009four}.
Additionally, insights on social behaviors require collective adoption from multiple socially related participants who must also consistently maintain the devices (e.g., keeping devices charged), thereby posing challenges to large-scale sensing and practical deployment of sensors. } Together these factors challenge the scalability of such methods because they provide a sparse representation of the community. 

An approach \edit{that mitigates some of the client-side challenges,} is to use infrastructure-based techniques, such as installing Bluetooth beacons into the built environment~\cite{eagle2006reality, DasSwain2019FitRoutine}. 
\edit{Nevertheless, these techniques can also rely on data being collected and processed through a client~\cite{hong2016socialprobe,das2018groupsense}.}
\edit{Moreover, augmenting the entire infrastructure with new sensors for comprehensive coverage can be a significant ask for many campuses. 
In addition, novel deployments cannot be used to inspect prior campus-scale social behaviors (e.g., exam week, violent incidents, shutdowns, and global infectious disease-related pandemics).
}
\edit{In contrast, many campuses maintain a managed WiFi access-point (AP) network that provides device association logs which can be repurposed to infer locations of users~\cite{shi2016walk} and subsequently model individual behaviors~\cite{ware2018large,eldaw2018presence}.
Albeit a coarse descriptor of location --- with low spatio-temporal resolution --- these WiFi association logs can describe collocation of individuals. \textit{Positing that these collocation behaviors \revision{present avenues for} social interactions, in this paper, we examine their relationship to \revision{the performance of students in project groups}
by 
harnessing data from a coarse sensor available in most modern campuses --- a managed WiFi network.} } Specifically, we pursue the following research goal: 
\revision{\textbf{To what extent is WiFi based coarse collocation associated with group members' academic performance?}}


\edit{It can be argued that AP logs are a naive means of localization because of the imprecise and variable coverage~\cite{kjaergaard2010indoor} along with irregular log updates~\cite{stallings1998snmp}. Meanwhile, APs are also ubiquitous, their data logging is consistent and managed networks typically archive multiple devices associations over a long period of time.} \edit{Accordingly towards our research goal, the paper first provides a system description that elaborates how we determine collocation from association logs.} 
\revision{Consequently, despite its low spatio-temporal resolution, we explore if unobtrusively inferred collocation of project group members is related to performance in the project.}

We note that, in itself, the collocation of individuals does not warrant that their actions are interlinked~\cite{rummel1976understanding}. Yet, when individuals gather in a space with a common intent, it can describe their relationship to each other. In this work, we  harness this aspect of human interactivity known as \textit{spatiality}~\cite{olson2000distance}, by studying 
\edit{
collocation behaviors of a set of students that are known to share their situated on-campus experiences}. 
\edit{These students were distributed across 54 course project groups in a single course, and  interacted over a 14-week period. We validate our approach by examining, using statistical modeling approaches, if a student's collocation patterns are associated with an established outcome of social interactions---performance in teams ~\cite{finch1997factor,ford1990relationship}.} 

\edit{
Essentially, this paper presents a case study that illustrates how collocation information, determined from campus WiFi association logs,} 
\revision{ can indicate a known outcome of presence in the same space --- improved performance ~\cite{geister2006effects,fruchter2010tension,mark2005no,kozlowski1987exploration,edmondson1999psychological}.}
In the light of our approach and findings, the paper highlights other domains 
where such coarse collocation data could be useful, such as mental well-being, and physical health. Our discussion additionally elaborates the privacy and data-ethics concerns related to practical deployments of such technology. We conclude by discussing the limitations of the work along with opportunities to extend this work in future studies.

\section{Background and Related Work}
\revision{Collocation enables members of a group to socially interact fluidly through both implicit and explicit actions~\cite{olson2000distance}. Particularly, when individuals with a common intent are in the same space at the same time, and are aware of it, they engage in some form of \textit{collocated synchronous interactions}.}
Note, this paper adopts the definition for social interactions as described by \citeauthor{rummel1976understanding}~\cite{rummel1976understanding}: \textit{``...acts, actions, or practices of two or more people mutually oriented towards each other's selves...''}. 
Rummel's definition refers to people regulating their actions based on others sharing mutual intent, with the purpose of shaping their subjective experience~\cite{rummel1976understanding}. Although these interactions can take place digitally, this paper focuses on \revision{automatically identifying} synchronous social interactions in the physical world, i.e., the people interacting are 
\revision{collocated.}
\vspace{-10pt}

\revision{\subsection{Collocation and Performance}\label{sec:rw-collocatio-performance}}

\revision{The CSCW and Organizational Psychology community has always been interested in understanding the value of collocation for work~\cite{olson2000distance,mark2005no,kozlowski2006enhancing,hinds2003out,geister2006effects}, especially to inform better designs for remote work technologies. Literature on collocation describes the importance of intense interlinked activities in a dedicated physical space~\cite{olson2000distance,kozlowski2006enhancing} (e.g., ``warrooms'') as well as fluid activities in the presence of coworkers in a general physical space~\cite{kozlowski1987exploration,mark2005no} (e.g., open offices or adjacent cubicles). Both forms foster social interactions that are associated with individual and team performance~\cite{kozlowski2006enhancing,edmondson2001disrupted, hinds2003out,geister2006effects,fruchter2010tension}. Therefore, exploring methods to identify and understand collocations can help communities design infrastructure and policy to support better performance.}

\revision{~\citeauthor{olson2000distance} characterize multiple aspects of collocation at work and its implications~\cite{olson2000distance}. Foremost, it is a synchronous social interaction that is not limited to verbal discussions and active sharing of resources. Even the presence of others working towards a common goal allows for subtle exchange of information through gestures and expressions~\cite{olson2000distance} (e.g, is a teammate struggling, are they too absorbed or are they available for feedback). Additionally, collocation provides shared context that comprises common points of reference (e.g., whiteboards, post-it notes, or verbal concepts)~\cite{olson2000distance}. Moreover, it supports informal interactions that can help ``opportunistic information exchange'' and improve social ties with teammates~\cite{olson2000distance}.}

\revision{Prior work also posits several reasons that link collocation to performance. Being physically situated in the same space keeps team members up-to-date, and therefore agile and innovative~\cite{kozlowski1987exploration}. Staying collocated helps maintain common mental models of tasks, resources, skills, and problems~\cite{cannondefining}. In contrast, distance is known to elicit more conflict~\cite{hinds2003out}. This is likely due to the non-uniform distribution of information that can lead to excluded members partaking in incomplete, inaccurate, or redundant tasks~\cite{cramton2001mutual}. Distributed work is also related to heightened tensions between teammates, which affect wellbeing and impede individual performance~\cite{fruchter2010tension}. On the other hand, collocation allows \textit{team learning}, where members feel ``safe'' to seek feedback, experiment, and resolve errors~\cite{edmondson2001disrupted}. Feedback from teammates is known to augment individual performance~\cite{geister2006effects}. Moreover, collocation can improve social ties between members~\cite{trainer2016hackathon} and therefore improve performance~\cite{sparrowe2001social}. Related to performance, ~\citeauthor{mark2005no} observe that the subtle cues of collocated social interactions are related to individuals focusing on single tasks for longer, continuous periods~\cite{mark2005no}. }

\revision{Traditional methods of evaluating collocated social interactions rely on survey instruments, but these are limited by recall and desirability biases~\cite{aiken1990invalidity,krumpal2013determinants}. Moreover, self-reports are static assessments,  while social interactions are fluid and vary over time~\cite{schroder2016modeling}. One approach to studying human phenomena by avoiding such biases is with unobtrusive sensing. These automatic methods have the promise of dynamically sensing human behavior without interfering with an individual's natural functioning and are, therefore, more practical for gathering reliable insights.}

\revision{\subsection{Automatically Sensing Collocation}}


Automatically sensing \revision{collocation} has interested the community for over a decade. 
Prior work in this space has fundamentally focused on two approaches, separated by the scale of collocation.

The first set of approaches have focused on studying face-to-face interactions \edit{and proximity} in small spaces~\cite{eagle2006reality,olguin2008sensible}. ~\citeauthor{olguin2008sensible}, used wearable badges to study how low-level interpersonal interactions are related to workplace performance~\cite{olguin2008sensible}. Although this methodology is precise, it is limited by the cost of instrumentation and by the obtrusiveness of wearing a foreign device. A known variation
of this is to use
Bluetooth sensors embedded in one's smartphone~\cite{wang2015smartgpa, meng2014analyzing}. 
However, 
installing applications on an existing wearable~\cite{tsubouchi2013working,hung2013classifying} or phone ~\cite{das2018groupsense} still requires user adoption, raises privacy concerns~\cite{shilton2009four}, and \edit{is constrained by manufacturer protocols to acquire this data}. 
The second approach has largely focused on studying city-scale ``flocks'' through GPS based localization~\cite{eagle2006reality,wang2016crowdwatch,fan2015citymomentum,kanhere2013participatory}. 
While this approach scales, \edit{practical deployments still require acquiring this data from client devices. Capturing GPS data from the infrastructure is a non-trivial task that many researchers and stakeholders will not be able to perform.}

\revision{
Inferring collocation of multiple socially related individuals requires approaches that can characterize long-term behaviors of groups.}
\revision{In fact, according to \citeauthor{lukowicz2011context}, one of the opportunities that describes \textit{socially aware computing} is ``methods for monitoring and analyzing social
interactions—in particular, with respect to long-term interactions and interactions within large communities and organizations''~\cite{lukowicz2011context}.}
As a result, campuses can consider harnessing data already logged through their network infrastructure without needing any active participant effort or involvement.


\vspace{-10pt}

~\revision{\subsection{WiFi-based Sensing of Collocation}}

\edit{In prior work, researchers have tried to determine \revision{collocation} through sensors in the environment that infer proximity.}
For instance, \citeauthor{hong2016socialprobe} have shown that WiFi-based fingerprinting can help identify ties between groups \cite{hong2016socialprobe}. \edit{However, to get comprehensive insights, such work still relies on robust clients and training the entire network }. 
\revision{Alternatively, enterprises have used WiFi router networks to develop Real-Time Location Systems (RTLS)~\cite{ciscoloc,accuwareloc}.}
To infer location, these technologies store the Received Signal Strength Indicator (RSSI) values for any client-device within a neighborhood of Access Points (APs). \revision{This could be extended to infer collocation but these solutions have a substantial cost for installation (requiring a full fingerprinting survey of the network). This coupled with the privacy concerns of excessive precision often outweighs the benefits of any realistic campus use-case.}
Yet, a common form of WiFi infrastructure deployment in university campuses~\cite{eldaw2018presence,ware2018large} only stores association logs describing which AP a client-device is connected to. Although it is relatively coarse~\cite{martani2012enernet}, this parsimonious representation of location has been exploited to understand individual behavior. ~\citeauthor{ware2018large} have inferred student location based on network logs to assist depression screening by inferring individual dwelling patterns 
~\cite{ware2018large}.
Similarly, ~\citeauthor{eldaw2018presence} have used unsupervised methods on similar logs to understand the relationship between student visits patterns and  
the semantic purpose of certain campus spaces~\cite{eldaw2018presence}. 
\revision{These works motivate us to approaches that adhere to data minimization.}
While prior examples trace individual dwelling patterns across campus, they do not explicitly assess group behaviors. 
\revision{We extend on such efforts to identify collocation between multiple students with a shared intent, such as a group project. }


Even though collocation does not necessitate verbal communication in the strict sense, it does serve a social function~\cite{olson2000distance} \revision{that is associated with the performance of collocated individuals (Section~\ref{sec:rw-collocatio-performance})}. 
Therefore, we seek to determine if 
\revision{automatically determined collocation patterns are related to performance of members of group projects, and can thus approximate collocated social interactions related to work.}
For instance, prior works have shown that mining WiFi network data can cluster people into social and behavioral groups~\cite{jiang2015mining,wang2018understanding}.  Even other infrastructure-based coarse location technologies, such as Bluetooth, have been used to capture subtle social interactions like synchrony within-group routines~\cite{DasSwain2019FitRoutine}. While these studies implicitly associate individuals together (e.g., distinguish students by dining hall), they do not  explore \revision{collocation} in physical spaces sufficiently. 
A more direct insight of collocation was demonstrated by~\citeauthor{zakaria2019stressmon}, who also leveraged 
the campus network infrastructure to 
predict stress~\cite{zakaria2019stressmon}. 
However, these systems either rely on additional augmentation of the infrastructure or knowledge of the network signal strength received by clients. In contrast, this paper explores if insights on human behaviors are evident in coarse collocation derived from rudimentary raw network association logs that could be applied to almost any managed wireless network today.
 

\vspace{10pt}

\section{System Description: Identifying Collocation with Network Logs}
\label{sec:rq1}


\begin{figure}[t!]
    \centering
    \includegraphics[width=\columnwidth]{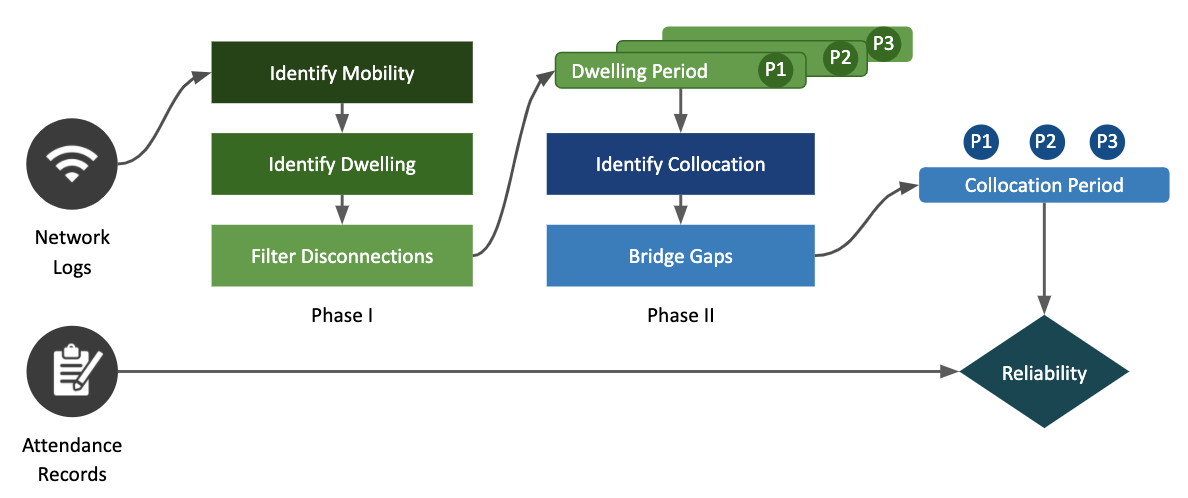}
  	\caption{Deriving collocation periods from raw WiFi network logs and compare this to attendance records}
  	   \label{fig:rq1_structure}
\end{figure}


A ``social interaction'' occurs when and where individuals mutually orient themselves~\cite{rummel1976understanding}. 
\revision{Collocation of individuals in the same physical space creates opportunities for social interactions where their behaviors and perceptions are interlinked~\cite{olson2000distance}}
In the scope of this paper, periods when individuals are collocated is the basic unit at which we infer social interactions. 
\revision{This study was done retrospectively to ensure students do not alter behavior under observation and to ensure their privacy is not compromised while they are enrolled in the course. Therefore, we use class attendance records to validate the system, as these can be obtained retroactively.} As illustrated in~\autoref{fig:rq1_structure}, this section describes a pipeline to determine collocation by leveraging WiFi network association logs and an evaluation of its reliability \revision{by comparing it to class attendance records.}

\subsection{Network Data}


To build a reliable processing pipeline we need real data that represents on-campus students. This is to ground the methodology in how the network logs actually depict the  behaviors of real students.

\subsubsection{Sample Association Logs}
We obtained consent from 46 students at a large public university in the United States, and then analyzed their anonymized WiFi association logs. These students belonged to two sections of a project-intensive course. Both sections were taught by the same instructor and had attendance data for each lecture. We refer to these sections as ``1A'' (22 students) and ``1B'' (24 students) throughout the paper. The instructor for the course provided each consenting student's attendance and group label, along with the course lecture schedule. 
\revision{We partner with the institute’s IT management facility to obtain network log data for these students. Through this collaboration we were provided logs associated with any device owned by a consenting student without requiring direct access to device MAC addresses.}
This data was accessed at the end of the semester\footnote{This analysis was approved by the Institutional Review Board (IRB) of the relevant institution, and the data was de-identified and secured in approved servers} and contains approximately 14 weeks of data, which spans 34 lectures for each section\footnote{No lectures took place 21st January 2019 (MLK Day), 1st week of January (winter break), and 3rd week of March (spring break)}.

\begin{wraptable}{r}{0.3\columnwidth}
\sffamily
\centering
\small
\setlength{\tabcolsep}{2pt}
\caption{Sample raw log entry
}
  \begin{tabular}{l|l}
    \toprule
    Field & Sample \\
    \midrule
    Timestamp  & Apr 1 00:10:51  \\   
    Update Type  & snmpupdate  \\  
    Anon. User  & 2099  \\  
    User Device  & c4:7d:eb:0f:df:d5  \\ 
    AP ID  & 40:cd:14:b2:02:c0  \\
    AP Label  & 122S-209  \\        

\bottomrule
\end{tabular}
\label{table:log_entry}
\end{wraptable}

\subsubsection{Managed WiFi Network}
\label{sec:pipeline_netdata}

Every AP installed on campus is mapped to a building ID and a room ID. The room ID indicates the room closest to the AP or the room that contains the AP. For instance,~\autoref{table:log_entry} shows an AP in room 209 of building 122S. Larger rooms, such as lecture halls, have multiple APs to increase coverage. In the logs, these APs are registered with different MAC addresses but associated to the same room.
Every entry in the log documents an SNMP (Simple Network Management Protocol) update in the network. \edit{This update is triggered when APs see a change, i.e., a device connects, or through an SNMP poll request to the AP that returns connected devices.}
Therefore, the log itself indicates that a device is in the vicinity of an AP, but without information of the client RSSI, this inference has a low spatial resolution.
\edit{Moreover, the logs for a connected device are }
erratic because of variable connectivity settings in the device agent (e.g., the WiFi turns off when inactive). The irregularity in log updates leads to a low temporal resolution. The low resolution is what introduces ``coarseness'' to this data. 
Outside of the specific association timestamps---when an AP responds to an SNMP poll or a client switches APs---the connected device is invisible in the logs. 








\begin{figure}[t!]
    \centering
    \includegraphics[width=\columnwidth]{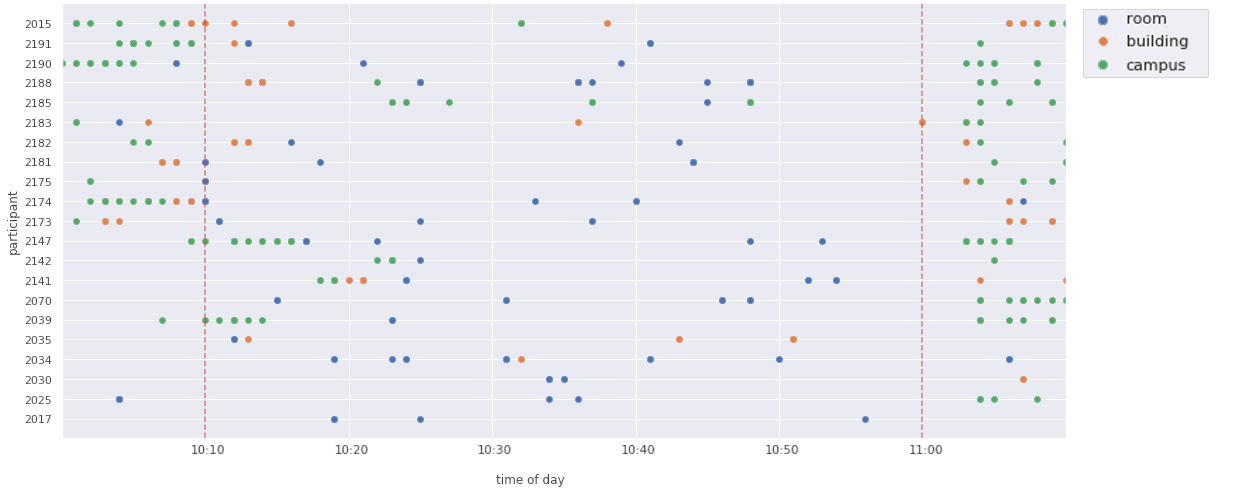}
    \vspace{-0.1in}
  	\caption{Each marker represents an SNMP update log for that participant, and the vertical red dashed lines indicate lecture time for section 1A on 5th April, 2019}
  	   \label{fig:lecture_raw}
  	  \vspace{-0.2in}
\end{figure}

\subsection{Phase I: Identifying Dwelling Segments from Raw Logs}


\edit{Since the logs are coarse, their raw form can only  describe device location at the moment it is timestamped. Therefore,  we first need }
to reliably determine where an individual is dwelling
\edit{between two successive log timestamps.}
To determine this, we applied the following process:



\textbf{\textit{(i) Determine if an Individual is Mobile --- }} We know the scheduled class time and location for the regular lectures of sections 1A and 1B. To assess how students move, we examine the logs accumulated in the $30$ minutes before and after the lecture. \revision{One of the classrooms had only 1 AP while the other one had 3 APs for coverage}. For the class held  on April 5, 2019, for Section 1A, Figure~\ref{fig:lecture_raw} depicts the instances when a student's device is logged before, during, and after the lecture times, along with the AP information for that log entry.  Only less than 1\% of the log entries show concurrent updates at different APs from two or more devices owned by the same student. This is why we treat all log entries from a student's device as a proxy for the student.  Since SNMP updates occur when a device roams, we measure the interval between two successive log entries from a user's device that associate with different APs. \revision{For example, from entering the building to entering class, devices will snap to different APs. This leads to 2 successive log entries at different locations. However, 2 such  entries do not necessitate the time between them was spent moving. Consider Participant 2173 in ~\ref{fig:lecture_raw}, who associates with an AP outside the building, then logs an entry at an AP in the same building before logging an entry in classroom entry, almost 8 minutes later. While it is possible that no AP was  move, but it is also possible the student was dwelling in an adjacent area and then moved to class when it started.} However, Figure~\ref{fig:lecture_raw} also illustrates that for most students the log updates before and after class times also exhibit higher update frequency in shorter intervals. As a result, we determine the  90th quantile of the  intervals between 2 different logs \revision{to learn a reasonable threshold that can capture devices roaming before the student settles into class.} This was found to be 233 seconds. \revision{Therefore, we consider devices moving when different APs successively log the same student's device below this threshold. } 
    
\begin{figure}[t!]
    \centering
    \includegraphics[width=\columnwidth]{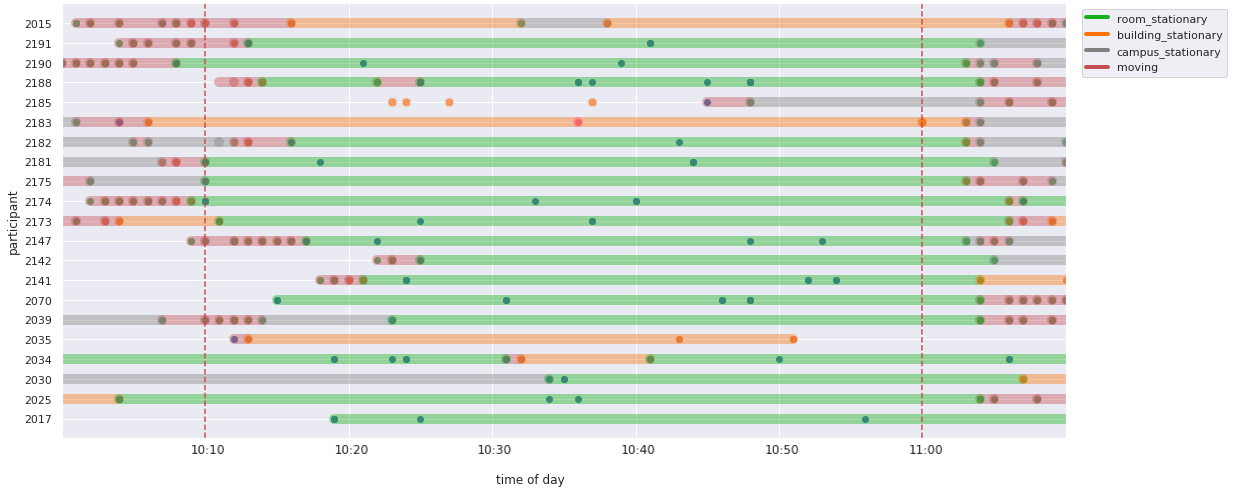}
    \vspace{-0.1in}
  	\caption{Once the moving segments of an individual has been identified, the time periods between these segments are interpolated as dwelling segments}
  	   \label{fig:lecture_segments}
\end{figure}

\textbf{\textit{(ii) Determine if an Individual is Dwelling in Place --- }} 
The user is considered to be dwelling at the location associated with an AP for  any time segment when they are not mobile. Based on the criteria for moving, a user is considered stationary \revision{in 2 cases, (i) when successive log entries are at the same location, (ii) the time before the next entry exceeds the threshold.} 
Contiguous dwelling segments where the AP does not change are combined to  represent longer dwelling segments. Figure~\ref{fig:lecture_segments} shows how the raw logs represented in Figure~\ref{fig:lecture_raw} can depict moving (red) and stationary (orange or green) time segments.


\begin{wrapfigure}{r}{0.35\textwidth}
  \begin{center}
    \includegraphics[width=0.35\textwidth]{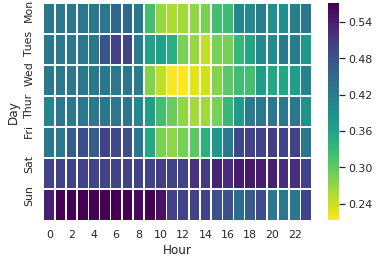}
  \end{center}
  \caption{The median portion of time a user is disconnected from campus for a given hour for a day of the week.}
    \label{fig:discon}
\end{wrapfigure}
\textbf{\textit{(iii) Filtering Out Disconnection Periods --- } }
\revision{When students exit campus they might disconnect from the network because of poor AP coverage outdoors.}
\revision{Individuals can be lost to the network and then be ``visible'' when they enter a building} after a period  of time. Due to our threshold, the time period between these two mobility phases can be erroneously labeled as dwelling, whereas the user was actually disconnected from the network. 
This large interval needs to be distinguished from actual dwelling periods. Based on the class dwelling time, we  find that the longest interval between two successive log entries of a student actually present in class was 76 minutes. We use this as a heuristic threshold. With this, we mark any periods of dwelling as disconnected (or inactive) where the log entries are timestamped at intervals exceeding the threshold. \autoref{fig:discon} shows that the disconnection periods identified were predominantly on weekends and before or after class times.


\revision{It is important to note that our work centers on collocation indoors. These same heuristics cannot be transferred to identify collocation outdoors as the AP network is not as dense outdoors. Such research would require a different set of heuristics outside the scope of our paper.}


\subsection{Phase II: Identifying Collocation}
\label{sec:pipeline_colloc}
Phase I identifies  dwelling periods for individuals. Phase II identifies overlapping dwelling near the same AP (or room) to describe collocation.
Simply considering the overlapping dwelling segments could have breaks when even one of the collocated members inadvertently switches between AP and then returns (e.g., participant 2034 in Figure~\ref{fig:lecture_segments}). 
This could occur either when they took a break or if they are in place but their device intermittently found a better connection to a different AP.
Since the aim of obtaining collocation segments, 
\revision{is to use it as a proxy for collocated social interactions (Sec~\ref{sec:rw-collocatio-performance}), }
we consider a liberal approach to characterize collocation. This decision aligns with~\citeauthor{rummel1976understanding}'s definition of social interactions, that just because an individual is not in sight, it does not signify the conclusion of social interactions~\cite{rummel1976understanding}.
For example, when an individual takes a brief break from a meeting to grab coffee or use the restroom. 
Therefore, instead of dissecting the collocation around such short-lived absences, these gaps in the segments are bridged. In particular, these gaps are characterized by (i) common members of a group are collocated before and after a gap; and (ii) during the gap some subset of members are still dwelling or collocated.
After identifying such overlapping segments, we first find the median duration of these gaps. The median in our data for such occurrences was 11m 7s. Any gaps less than this threshold are resolved by considering all members to be collocated throughout, including the break period.


\begin{figure}[t!]
    \centering
    \includegraphics[width=\columnwidth]{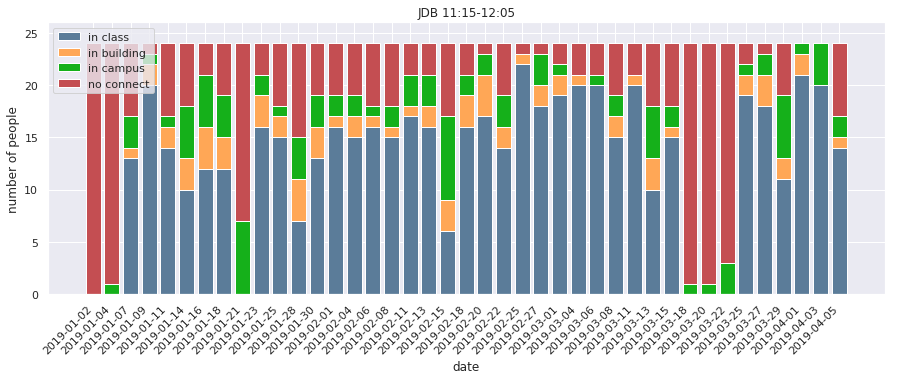}
    \vspace{-0.1in}
  	\caption{Each stack depicts where how many students of Section 1B were found to be connected to the lecture room's AP, another AP in the same building, to the campus network, or not connected at all}
  	   \label{fig:lecture_attend}
    \vspace{-0.1in}
\end{figure}

\subsection{System Reliability}
\label{sec:attendance-validation}
To quantify the reliability of this coarse localization and collocation technique, we evaluate the attendance of 46 students in 2 sections for the 34 lectures that occurred in the sample data period. Each section had 3 classes a week and but met in different buildings.
For both sections, the instructor provided us with lecture-by-lecture records of each consenting student's attendance.
\revision{Attending class is one form of collocation on campus that involves students gathered around a WiFi AP. Even though every AP's coverage on campus might vary, when students do collocate to work outside lecture times they typically gather in breakout rooms, empty classrooms, library spaces, or other similar indoor spaces. Hence, we consider presence in class a reasonable }
ground truth to evaluate the reliability of our proposed automated method \revision{for the purposes of our study}.




\textbf{Missing Data. }
First, we would like to address the missing data problem. On certain lecture days, we did not find any entry for some students (including a 30 minute margin before or after). The red stacks in \autoref{fig:lecture_attend} show the number of students per lecture with no log entries for section 1B. On comparing this to the attendance records, we learn that 93\% of the times a student does not appear in the logs, they were actually recorded as present by the instructor. One possibility is that the student either had all their devices turned off or connected to a different network (e.g., cellular data, \edit{ or the campus guest/visitor network}). 
Every student in our sample had no WiFi log entries on at least one lecture they attended (the median was five    lectures). Therefore, despite its pervasiveness, leveraging the managed network can still miss out on students who were actually present. 
For such occurrences, the automated method cannot ascertain presence or absence and therefore, we exclude these student records (for that lecture) from further analysis.






\begin{wrapfigure}{r}{0.3\columnwidth}
  \begin{center}
    \includegraphics[width=0.3\columnwidth]{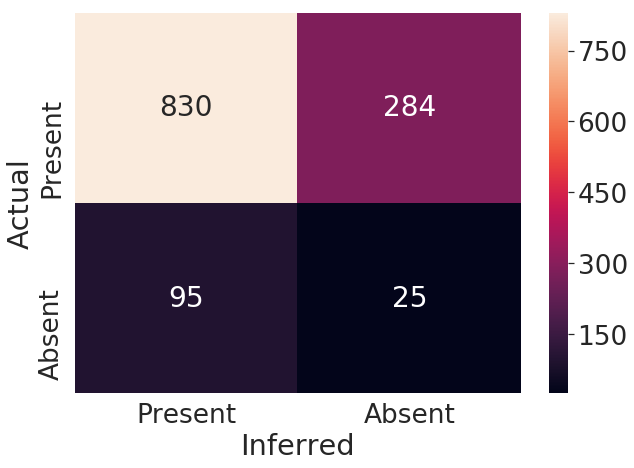}
  \end{center}
  \vspace{-1em}
  \caption{Actual vs inferred attendance; Precision: 0.89,  Recall:  0.75
  }
    \label{fig:conf_mat}
\end{wrapfigure}

\textbf{Accuracy. }
We consider a student to be in class if any time during class they were "seen" as connected to the AP associated with the room of the lecture.  They then show up as collocated with their peers in ~\autoref{fig:lecture_segments}, and their time in class is depicted by green segments.  We found $89$\% agreement between the instructor's record of who was present and our estimated record, a \textit{precision} measurement. \revision{Also, the ~\textit{false discovery rate} is $0.103$. Therefore, WiFi logs rarely indicate a student is at a location when they are not physically present.} We speculate, the false positives that emerge are because of students failing to record their name on the attendance sign-up sheet, possibly because of showing up late to class.

Alternatively, for every instance when the student was present, this method infers them to be collocated $75$\% of the time---\textit{recall}. \revision{For reference the \textit{false negative rate} is $0.25$}. Together, these indicate a relatively high proportion of false negatives ~\autoref{fig:conf_mat}. A false negative could occur when a student's device connects to a different AP on the network.~\autoref{fig:lecture_segments} denotes these as the orange segments. A device could also connect to an AP that is physically further away because the signal from their closest WiFi was attenuated~\cite{kjaergaard2012challenges}. Therefore, this uncertainty in location could still lead to missing out on students that were actually present. 

\revision{To summarize, the \textit{F1-score} of such a system can be interpreted as $0.81$. It has high precision, but with a \textit{specificity} of ($0.74$), it can erroneously mark students as absent when they were present.} In the future, this can be addressed by deploying a broader set of APs for a given location. 




\vspace{2em}
\section{Case Study: Collocation and Performance in Groups}
\label{sec:main-case-study}



\revision{
Our central motivation is to assess if 
coarse collocation patterns ---detected by repurposing network logs---
are related to  academic performance and thus approximate collocated social interactions~\cite{olson2000distance}}.
\revision{Collocation in teams is known to be associated with performance of workers.~\cite{olson2000distance,mark2005no,kozlowski2006enhancing,hinds2003out,geister2006effects}}
This encourages us to investigate the relationship between a group member's performance and how they collocate with other group members (such as time invested in meetings, the regularity of group activities, and the locations of these meetings). 
This case study demonstrates the feasibility of leveraging raw logs for one specific application that involves \revision{collocation}---understanding the performance of group project members.  In this way, it answers our research question,
\revision{\textit{To what extent is WiFi based coarse collocation associated with group members' academic performance?}}

\subsection{Study}

The participants were enrolled in an undergraduate design course for CS students. The course is offered every semester and is a two-semester sequence typically taken by students in their junior (3rd) year. 
Students in this course were expected to work with a team of four to six students over two semesters (Part 1 and Part 2) on a single design project.  In Spring 2019, this course had four sections for Part 1 and five sections for Part 2. Each section had an enrollment of about 40 students.
In terms of course structure, Part 1 involved both lectures as well as project milestones. In contrast, Part 2 had fewer lectures and expected students to allocate scheduled class-times for project-related efforts. Students in both parts were expected to collaborate on project work outside scheduled lectures. It is not generally known how often student teams met outside of class, nor is it known how much those collocations impacted performance. The data used in the previous section was from a subset of sections of this same course that also had attendance records (Section~\ref{sec:rq1}).


\subsubsection{Participants}
\label{sec:study_rec}

\begin{wraptable}{r}{0.3\columnwidth}
\sffamily
\centering
\small
\setlength{\tabcolsep}{6pt}
\caption{Participants in the study with complete data 
}
  \begin{tabular}{l|rr}
    \toprule
      Section & Part 1 & Part 2 \\
    \midrule
    A  &  22 & 21 \\ 
    B  &  24 & 27 \\   
    C  &  18 & 31 \\ 
    D  &  20 & 12 \\ 
    E  &  -  & 11 \\ 
    \midrule
    Total & 84 & 102 \\ 
    
\bottomrule
\end{tabular}
\label{table:pivot_part}
\end{wraptable}


\textbf{Recruitment. }
The recruitment took place in Spring 2019 in collaboration with the course instructors. The research team advertised the study during the lectures and online outreach through the instructors. In addition, a large number of the students were recruited during the final demonstration expo that is attended by students of both parts. 
Upon enrollment, participants provided consent for the researchers to access their anonymized WiFi AP log data as well as their course data. The participants were assured that this is retrospective data that is already archived and the insights of our study would not  impact their course outcomes. During enrollment, participants also completed an entry survey where they reported their group ID along with describing when, where, and how often they interacted with their group members face-to-face for class purposes. Participants were remunerated with a $\textdollar5$ gift-card for enrolling. 
In total, we received consent from 186 students (\autoref{table:pivot_part}). Of these, 170 students were in the age range of 18-24 years, and 16 were of age 25 and above. Among these students, 59 reported female (32\%)\footnote{As per the official headcount 25\% of the students within the CS major have been recorded as female}. 

\begin{wrapfigure}{r}{0.35\textwidth}
  \begin{center}
    \includegraphics[width=0.35\textwidth]{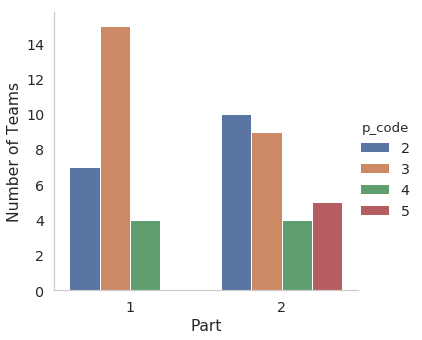}
  \end{center}
  \caption{Distribution of group sizes among the students recruited. At least one other member of their group must consent for a student to be included.}
    \label{fig:bar_grp}
\end{wrapfigure}

\textbf{Privacy.} Given the nature of data being requested, participant privacy was a key concern. The two core streams of data, course outcomes and WiFi AP logs, are both de-identified and stored in secured databases and servers which were physically located in the researchers' institute and had limited access privileges. 
The study and safeguards were approved by the Institutional Review Board of the authors' institution.

\subsubsection{Course Data}
After grading for the semester was completed, the different instructors of the course provided course-related data for 186 consenting students along with course lecture times (\autoref{table:pivot_part}). Among these students, 23 students did not have any other member from their group in our study and thus were dropped from this analysis. These remaining 163 students were in  54 separate groups (~\autoref{fig:bar_grp}). 


\textbf{Final Score.} All instructors provided the final score of  their section's students. This is a numerical score between 0 and 100 that informs the eventual letter grade based on the instructor's grading scheme. 
This final score is dominated by the project outcomes but students are assessed individually. These variations are introduced by participation as well as the instructor's subjective assessment of peer evaluation. \edit{Among the recruited group members, the range of scores between members could be as large as $6.5$ points.} 
This final score represents the ground truth for a student's academic performance.

\begin{wraptable}{r}{0.4\columnwidth}
\sffamily
\centering
\small
\setlength{\tabcolsep}{2pt}
\caption{Distribution of Peer-Evaluation Scores; all scales 1-5 except Psychological Safety (1-7) 
}
  \begin{tabular}{p{3cm}|rrr}
    \toprule
    Construct & Mean & Med & Std  \\
    \midrule
    Member Effectiveness  & 4.36 & 4.45 &  0.51  \\   
    Team Satisfaction  & 4.44 & 5.00 &  0.76  \\   
    Psychological Safety  & 6.12 & 6.29 & 0.80  \\ 
    Conflict (Task) & 1.64 & 1.67 &  0.62  \\ 
    Conflict (Relation) & 1.26 & 1.00 &  0.51  \\ 
    Conflict (Process)  & 1.41 & 1.00 &  0.59  \\ 

\bottomrule
\end{tabular}
\label{table:dist_peereval}
\end{wraptable}

\textbf{Peer Evaluation.} 
\label{sec:study_pe}
Given the group project nature of the course, 
students completed a fairly extensive peer-evaluation battery. This battery was completed by the students at the end of the semester and it captures their perceptions of conflict, satisfaction, and security with the team~\cite{jehn2001dynamic,van2001patterns,edmondson1999psychological}. It can also assess behaviors like collaboration, contribution, and feedback~\cite{loughry2007development}. In essence, this battery evaluates an individual's \revision{subjective experience while working in a team, their relationship with the team, and perception of other members}. Prior work shows that these instruments quantify aspects of social interactions that relate to performance~\cite{carnevale1998social, taylor1988illusion, jiang2013emotion, jehn1997qualitative}. Therefore, we use a participant's responses to these surveys as a gold-standard to infer their score. We feed these responses into a model to compare against the predictions of models trained on automatically inferred behaviors.
~\autoref{table:dist_peereval} summarizes the distribution of scores 
for each peer-evaluation survey instrument. The peer-evaluation  contained the following validated survey instruments:


\begin{itemize}
    \item \textit{Team Conflict~\cite{jehn2001dynamic} ---}  Conflict represents the perception of incompatible goals or beliefs between individuals that cannot be trivially reconciled. This battery contains three scales, ``task conflict'', ``process conflict'', and ``relationship conflict''. 
    \revision{When individuals perceive less conflict it is associated with performance enhancement~\cite{carnevale1998social, taylor1988illusion}. This is likely because the positive outlook leads to better motivation~\cite{taylor1988illusion} and satisfaction~\cite{jiang2013emotion}}.
    \item \textit{Team Satisfaction~\cite{van2001patterns} ---} Satisfaction reflects the contentment of an individual with their situation in terms of their expectations. Dissatisfaction with one's team can lead to lower levels of task performance~\cite{taylor1988illusion,jiang2013emotion} and also moderate the effects of conflict on performance~\cite{jehn1997qualitative}.
    \item \textit{Psychological Safety~\cite{edmondson1999psychological} ---} This construct captures a \textit{``shared belief held by members of a team that the team is safe for interpersonal risk taking ''}~\cite{edmondson1999psychological}. 
    This is associated with individual learning progress \revision{as they are more amicable to experiments and feedback~\cite{edmondson1999psychological}.}
    \item \textit{Team Member Effectiveness~\cite{loughry2007development} ---} This measure encompasses five dimensions~\footnote{While the other scales were self-evaluations, this score is the average of how their peers evaluated a team member}: (i) contributing to the project; (ii) interacting with collaborators; (iii) monitoring progress and providing feedback; (iv) expecting quality; and (v) relevant knowledge and skills. These characterize behaviors related to \revision{the individual-level construct,} ``team member effectiveness''~\cite{loughry2007development}.  
\end{itemize}




\begin{figure}[t!]
    \centering
    \includegraphics[width=0.9\columnwidth]{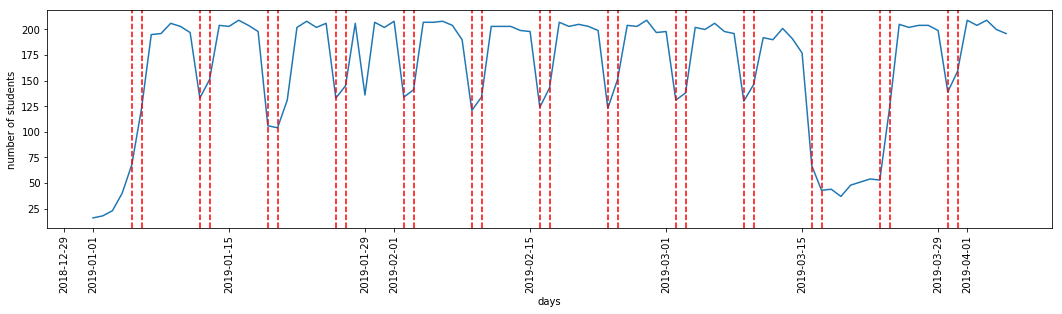}
  	\caption{The number of connected students reduces during the spring break (week of 15th March) and also coincides with weekends (depicted by vertical red dashed lines).}
  	   \label{fig:ping_semester}
  	  \vspace{-0.15in}
\end{figure}




\subsubsection{Network Data}
\label{sec:study_netdata}


The WiFi access point log data for consenting students was obtained from the institute's IT management facility. Since this data was already aggregated for maintenance and security purposes throughout the semester, we were able to retroactively obtain
this information at the end of the semester. The data spans all WiFi access logs by connected devices belonging to consenting students. This data is richer compared to the sample data for processing the raw logs into collocation (Section~\ref{sec:pipeline_netdata}). It includes more individuals and a larger set of APs.
The data spans a time frame of 95 days between January 1 2019 and April 5 2019. On average, the time between the first log entry for any one of a participant's devices and the last is approximately 90 days. 
\autoref{fig:ping_semester} shows the distribution of connected students throughout the semester. 
The logs in this study include 204 unique buildings with 4,865 unique APs. We also find multiple APs to be in the same room for 803 rooms.
Additionally, the 204 buildings were manually categorized to best express the purpose of that space~\cite{eldaw2018presence,wang2015smartgpa} --- for example, ``academic'', ``dining'', ``green spaces'', ``recreation'', and ``residential''. Two researchers referred to campus resources to independently assigned categories to these buildings.
Only two of the building labels disagreed,  
which was resolved by a third researcher. The raw logs of the consenting students was processed as described in Section~\ref{sec:pipeline_colloc} to obtain periods when students were dwelling and collocated. Through the semester, the median collocation duration of a student with another was about 70hrs.




\subsection{Feature Engineering}

The low spatial resolution of the collocation makes it insufficient to assert from 
\edit{isolated instances if collocation of group members was was connected to their performance.}
However, processing multiple collocation periods over the semester can represent behaviors  
\revision{that approximate collocated interactions relevant to performance (such as motivation, feedback, conflict resolution, opportunistic discussions, etc.)}
For instance, members of the same group might collocate regularly at a specific type of building. Therefore, we important  engineer features that can capture such patterns. 





\subsubsection{Feature Extraction}
This phase extracts relevant information at a week-level based on various behaviors labelled semantically with the help of manually annotated or retrieved data (e.g., building categories, group meeting/lecture schedules). We segregate features  by ``individual'' and ``group'' to capture different behavioral signals. The former is meant to characterize individual behaviors which are not explicitly \revision{social, but could impact performance (e.g, attendance.)}. The latter captures the behaviors of individuals that are oriented towards their group, such as time spent collocated with other group members. The dissociation between these features is meant to distinguish the explanatory power of the collocation behaviors from individual ones. This helps provide discriminant validity and assert that coarse collocation-based features are not confounded by an individual's general behavior, such as the time spent in academic spaces.~\autoref{table:feat_disc} summarizes the different features we extract at a week level. We derive the individual features based on the lecture schedule and semantic labels for buildings. To craft the collocation features, we use the same information but compute them as both absolute duration and a relative percentage. The former denotes how much time a student spent collocated with their group (at least one other member). The latter describes this behavior relative to the total time spent by that group together to express what portion of time a student 
\revision{was possibly included in a group's offline synchronous presence.}

The collocation features are crafted to consider when the behavior occurred:
\begin{enumerate}
\item \textit{Scheduled:} Groups reported their regular meetings in a free-form response field during enrollment (Section~\ref{sec:study_rec}). The meeting locations reported were at a building resolution and respondents typically indicated a primary building (e.g., learning commons) along with a potential backup (e.g., library). However, teams also expressed meetings could take place at undetermined locations on campus. Moreover, groups often provided multiple tentative meeting times and places for a week. To accommodate all possibilities, this feature captures the collocations between group members that occurred during any of the reported periods. 
\item \textit{Class:} This segregates collocations with group members during class times. This distinguishes itself from the attendance feature by considering periods of collocation even outside the assigned lecture room. For instance, in the case of Part 2 sections, the students were expected to meet among themselves during class time, and not necessarily in the scheduled room for the class. Based on student reports, Part 2 teams did not necessarily use all class times in a week for meetings. This feature represents this set of behaviors.
\item \textit{Other:} This is a catch-all bucket to capture all other ad-hoc collocations. Only 4 groups in our study reported interacting with group members for non-academic reasons (e.g., ``lived together''). \revision{Since, improvement of social bonds is related to performance~\cite{olson2000distance,trainer2016hackathon}}, this category encompasses impromptu \revision{collocations, that could be motivated by course milestones but also represent other serendipitous  situations}.
\end{enumerate}

\begin{table}[t]
\begin{minipage}[t!]{\columnwidth}%
\sffamily
\centering
\small
	\setlength{\tabcolsep}{5pt}
	\caption{Description of the raw features derived from the collocation data at a weekly level
}
  \begin{tabular}{l|p{4cm}|cccc}
    \toprule
    Type & Description & \multicolumn{4}{c}{Spatial Variants}\\
    \midrule
        & & Any & Academic & Residential & Recreational\\
   \rowcollight \multicolumn{6}{l}{\textbf{Individual Features}}\\
   Attendance & Present at lecture room during scheduled time  & \iconnone & \iconcheck & \iconnone & \iconnone \\
   Dwell & Time spent at a place while stationary & \iconcheck & \iconcheck & \iconcheck & \iconcheck \\
    &      &  &  &  &  \\
    \rowcollight \multicolumn{6}{l}{\textbf{Collocation Features}---\textit{Measured as absolute duration and relative to the group}}\\
    Scheduled Collocation & Time spent with group members during reported weekly meeting times  & \iconcheck & \iconcheck & \iconcheck & \iconcheck \\
   Class Collocation & Time spent with group members during class hours & \iconnone & \iconcheck & \iconnone & \iconnone \\
   Other Collocation & Time spent with group members at other times & \iconcheck & \iconcheck & \iconcheck & \iconcheck \\
   \bottomrule
\end{tabular}
\vspace{-0.15in}
\label{table:feat_disc}
\end{minipage}\hfill
\end{table}

\subsubsection{Feature Processing}
Raw week-level features were aggregated to derive features that describe collocation behavior. All the raw features we extracted (\autoref{table:feat_disc}) from the data are computed at a week-level for 14 weeks---$5\times14$ for \textit{individual features} and $(9\times2)\times14$ for \textit{group features}. This leads to a rather large feature space given the target variable was the final score obtained at the end of the semester. Therefore, to reduce the feature space we calculate summary features to describe the entire semester of the individual. Specifically for each feature extracted at a week level, we compute the \textit{median}, the \textit{mean} and the \textit{standard deviation} for the study period. These are moment statistics that quantitatively depict the  distribution of that feature throughout the study period. In addition to these, we also compute the \textit{approximate entropy} of the feature per individual~\cite{pincus1991regularity}. This statistic is  a measure of the regularity of that feature for every individual. This reduces the overall feature count to $20$ and $72$ for individual and group features, respectively.



\subsection{Training and Estimation}
Collocation is known to be related to performance
~\cite{olson2000distance,mark2005no,kozlowski2006enhancing,hinds2003out,geister2006effects}.
To study this, we build multiple models to investigate how the collocation-based features estimate final scores in comparison to survey-based peer evaluation scores. Since the final score is a continuous value, we estimate it using regression. This analysis aims to demonstrate the
\revision{extent to which coarse inferences of collocation between group members is related to their final score (RQ).}

\subsubsection{Model Descriptions}
\label{sec:validation_models}
\Mpe{} denotes the model trained on peer-evaluation scores (Section~\ref{sec:study_pe}) based on the self-reported survey responses provided by the instructors. This model illustrates the efficacy of peer-evaluation reports in  describing performance and serves as a benchmark because these constructs have been validated to be associated with performance~\cite{jehn2001dynamic,loughry2007development,edmondson1999psychological,van2001patterns}. \Mind{} refers to the model trained on individual features and therefore is independent of the participant's group. \Mgro{} describes the model trained only 
features that represent 
collocation among group members and therefore potentially describing collocated social interactions.
By comparing these models to a specific subset of features (individual or group), it is possible to assess the discriminant validity in predicting final course scores with each subset without confounding interaction effects from other features. 
\edit{Furthermore, we develop mixed models to comprehensively understand how a combination of  features estimate academic performance. We do this to investigate if \revision{automatically generated} features contain complementary signals or whether some features are redundant in the presence of others.} For this, \revision{we consider a model which mixes collocation behaviors within the group and individual behaviors (\Mindxgro{}).}


\subsubsection{Estimators and Validation}
We evaluate all models through a 5-fold cross-validation process \edit{ensuring that members of the same project group  do not span separate folds}. To estimate the target variable (the final score), for each model described, we train with different estimators to account for variations in the data. Particularly, we train a Linear Regressor~\cite{seber2012linear} to represent linear relationships between features and a Decision Tree Regressor~\cite{safavian1991survey} for non-linear relationships. Additionally, we also train a Gradient Boost Regressor~\cite{friedman2001greedy}, i.e.,  an ensemble method and thus a more sophisticated learner.
To determine the relationship between model features and final scores, we measure the correlation between the predicted value and the actual values. 
\edit{Since correlation can capture the directionality and trend of scores, it is a more practical estimator for academic scores which are often graded relatively.}
For internal validation, we compare these models to a rudimentary baseline \Mz{}, which always estimates the median of the target variable from the training set. 

\subsubsection{Feature Transformations and Selection}
\begin{figure}[t!]
    \begin{minipage}[t!]{1\columnwidth}
      \begin{subfigure}[b]{0.33\columnwidth}
        \centering
      \includegraphics[width=\columnwidth]{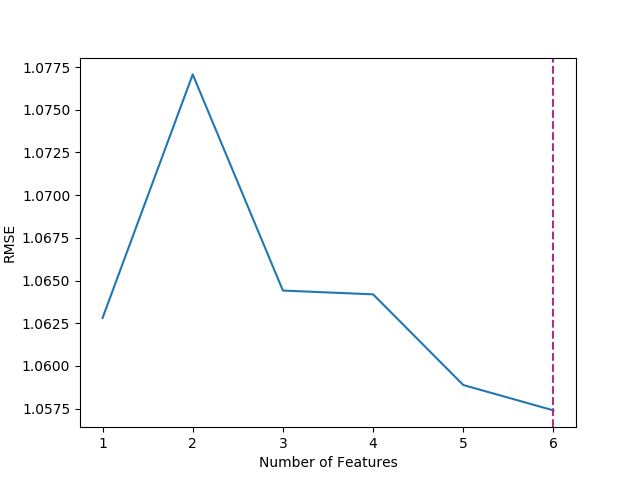}
      \caption{\Mpe{} with Linear Regression}
          \label{fig:feat_sel_pe}
        \end{subfigure}
      \begin{subfigure}[b]{0.33\columnwidth}
        \centering
      \includegraphics[width=\columnwidth]{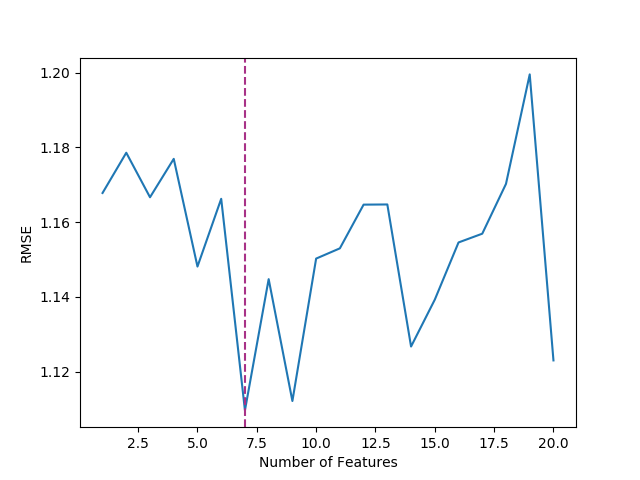}
      \caption{\Mind{} with Gradient Boost}
          \label{fig:feat_sel_iwf}
        \end{subfigure}
      \begin{subfigure}[b]{0.33\columnwidth}
        \centering
      \includegraphics[width=\columnwidth]{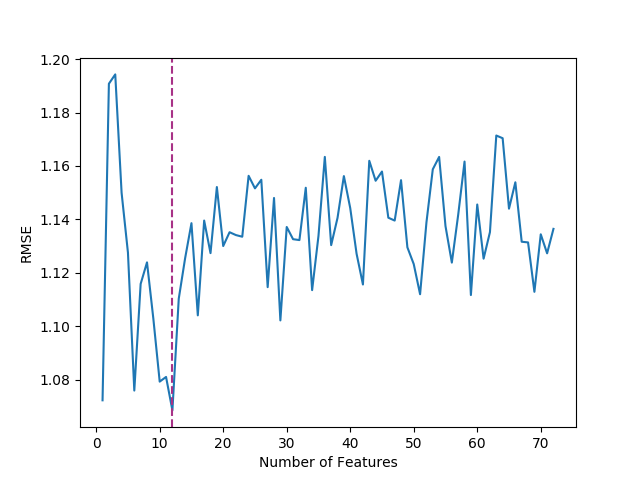}
      \caption{\Mgro{} with Gradient Boosting}
          \label{fig:feat_sel_qwf}
        \end{subfigure}
        \caption{Best number of features (X-axis) based on minimizing RMSE (Y-axis) with mutual information }
        \label{fig:feat_sel_ml}
    \end{minipage}\hfill
\end{figure}

The transformations are needed to solve problems with missing data and to scale the features to comparable units. The transformations and selections take place within each fold and therefore we perform these only with the training data of that fold:

 
\begin{enumerate}
    \item \textit{Scaling Final Scores by Instructor ---} The target variable the models are trying to estimate is the final score for the course. Since the final score varies based on the instructor, we standardize the final scores based on the distribution of scores for each instructor in the training data. 
    \item \textit{Impute Missing Data ---} For a few individuals certain features might have missing values. 
    For instance, some 
    students had not have completed all survey instruments. For some of the collocation features a few project teams did report their scheduled meeting times (7 students). We impute these missing values with the mean of the feature (after scaling).
    \item \textit{Standardize the Features ---} We convert all  features  to zero mean and unit variance~\cite{kreyszig2010advanced}. 
    \item \textit{Mutual Information Regression ---} Lastly, we employ a univariate feature selection method on the basis of mutual information between the training features and the target variable~\cite{kraskov2004estimating}. The number of features selected varies from 1 to $k$, where $k$ is the total number of features in the model. We select the $k$ that minimizes the RMSE (Root Mean Square Error)~\cite{chai2014root}. The choice of $k$ is illustrated in Figure~\ref{fig:feat_sel_ml}.
\end{enumerate}
 
\subsection{Results}
\label{sec:validation_results}
These results aim to  
delineate if patterns in coarse collocation 
\revision{are associated with individual performance.}
This section compares
various models (described in Section~\ref{sec:validation_models}). 

\begin{figure}[t!]
    \begin{minipage}[t!]{1\columnwidth}
      \begin{subfigure}[b]{0.33\columnwidth}
        \centering
      \includegraphics[width=\columnwidth]{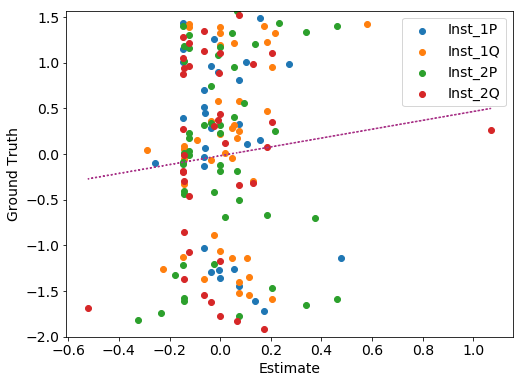}
       \caption{\Mpe{} with Linear Regression}
          \label{fig:res_pe}
        \end{subfigure}
      \begin{subfigure}[b]{0.33\columnwidth}
        \centering
      \includegraphics[width=\columnwidth]{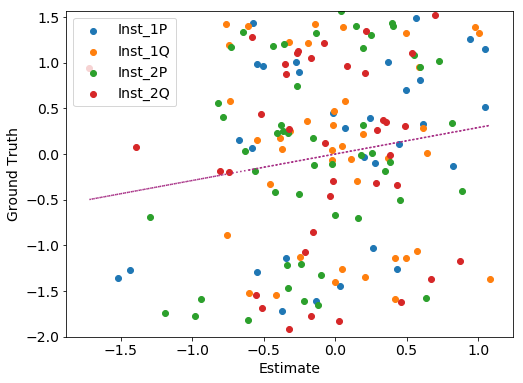}
      \caption{\Mind{} with Gradient Boost}
          \label{fig:res_iwf}
        \end{subfigure}
      \begin{subfigure}[b]{0.33\columnwidth}
        \centering
      \includegraphics[width=\columnwidth]{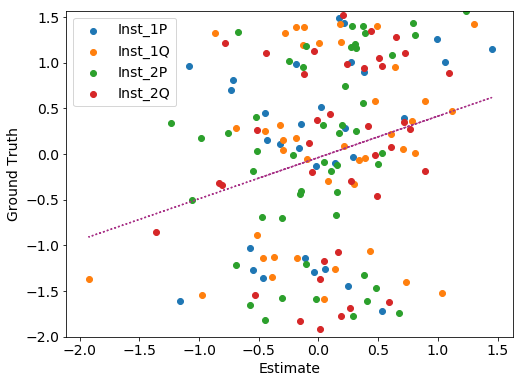}
      \caption{\Mgro{} with Gradient Boost}
          \label{fig:res_qwf}
        \end{subfigure}
        \caption{Comparing the different models in their estimation (X-axis) of an individual's final score (Y-axis); different instructors are labeled by different colours}
        \label{fig:res_ml}
    \end{minipage}\hfill
\end{figure}


\subsubsection{Model Comparison}
~\autoref{table:res_sum} summarizes the results
with the best estimator for each model. 
For any set of features, only the estimator that minimizes the RMSE is considered for comparison between models.\revision{To compare models we use \textit{Pearson's r} to describe the covariance of each model's estimate with the final scores of the students. This coefficient characterizes the complete association by considering all observations and does not assume normality~\cite{nefzger1957needless}.
}
\edit{All models exhibited an improvement over \Mz{} --- the rudimentary median estimator.}
\edit{None of the models based on peer evaluation features (\Mpe{}) were found to be significant, but among them Linear Regression showed the most error reduction. For \Mind{} the best estimator used Gradient Boost. Its estimates were more significant but with a weak correlation of $0.14$.}
\edit{In comparison, for \Mgro{} the best estimator, which used Gradient Boost, exhibited a very significant correlation of $0.24$. 
\revision{We also compare the dependent overlapping correlations of \Mgro{} against \Mpe{} and \Mind{}, by using the approach proposed by ~\citeauthor{zou2007toward}~\cite{zou2007toward} (with a confidence-interval of $90\%$). In both cases, the correlation of \Mgro{} with the final score is significantly different than that of \Mpe{} ($p = 0.02$) and \Mind{} ($p = 0.08$).}} Additionally, incorporating both individual and within group behaviors shows minor improvement. \revision{This improvement was not significant in comparison to \Mgro{}~\cite{zou2007toward}. }
Figure~\ref{fig:res_ml} shows the correlation \revision{coefficients of different models.}

\begin{table}[t!]
\begin{minipage}[t!]{\columnwidth}%
\sffamily
\centering
\small
	\setlength{\tabcolsep}{12pt}
\caption{\edit{Summary of Model Performance. (`-':p<1, `.':p<0.1, `*':p<0.05, `**':p<0.01)} 
}
  \begin{tabular}{llc|rl}
    \toprule
    Model & Training Data & Estimator & \textit{Pearson's R} \\
    \midrule
    \Mpe{} & Peer Evaluation  & LR & 0.08 & -  \\   
    \Mind{} & Individual Behavior  & GB & 0.14 & . \\ 
    \Mgro{} & Collocation Behavior  & GB & 0.24 & ** \\ 
    \Mindxgro{} & Individual + Collocation  & GB & 0.25 & ** \\ 

\bottomrule
\end{tabular}
\label{table:res_sum}
\end{minipage}\hfill
\end{table}


\subsection{Interpretation of Results} 
The results show that the model trained on students' collocation behaviors (\Mgro{}) outperforms the correlation of estimates obtained by modeling peer-evaluation and individual behaviors.
First, we find the collocation-based behaviors are significantly better at estimating final scores than peer-evaluation scores (\Mpe{}). While peer evaluation scores are expected to yield better correlations~\cite{taylor1988illusion, jiang2013emotion, edmondson2001disrupted, jehn1997qualitative}, the social desirability bias in manually reporting team experiences can wash out the intricacies of actual team behavior~\cite{aiken1990invalidity,krumpal2013determinants}. These surveys expect the participants to subjectively interpret and reduce their social experience into scores. But these students are also aware that these scores might affect the instructor's impression of their team members and possibly their own score. In contrast, \Mgro{} incorporated multiple characteristics of the collocation behavior within groups over multiple weeks. These features are devoid of the subjective biases that plague self-report and other manual assessments of collocated interaction. 
\revision{Note, this passive inference does not explicitly discern what transpired during collocation incidents. However, it can be a complementary source of data that describes student outcomes.} 

Second, we find that \Mgro{} performs better than a model built on individual behaviors (\Mind{}). Note that  \Mind{} was also found to be somewhat better than the peer-evaluation model. This already implies that dynamic offline behaviors 
\revision{have a significant relationship with } academic performance. However, given the collaboration-based nature of the course in determining the final score of an individual, \Mind{} falls short of \Mgro{}. This indicates that the individual behaviors of attendance or dwelling in academic spaces were not comparable in estimating the final score. This result indicates
\revision{that even in academic settings, collocation of students is important in courses that require agile coordination and collaborative work.}
\edit{This observation is in line with the concept of \textit{spatiality}, which describes that the presence of peers in the vicinity can affect individual performance even without direct communication~\cite{olson2000distance}. 
The features in \Mgro{} aggregate collocation behaviors of \revision{students known to be socially connected over}
multiple weeks. Particularly, participants in our group were expected to meet in person to work on their project towards their final score.}
\revision{Additionally, a very small proportion of students reported collocation with team members for reasons unrelated to their project.}
Therefore, the fact that the collocation based model (\Mgro{}) estimates the final score better than the dwelling-only model (\Mind{}) provides evidence that 
\revision{inferring collocation of socially related individuals help understand their performance.}
\edit{Moreover, \Mindxgro{}, which includes both group and individual behaviors, shows only a minor improvement over \Mgro{}. This further validates that it is indeed the collocation features that predominantly is related individual performance in settings like group design projects. }




\begin{wraptable}{r}{0.45\columnwidth}
\sffamily
\centering
\small
\setlength{\tabcolsep}{6pt}
\caption{Top 5 collocation features based on importance for \Mgro{} with Gradient Boost. The importance (Imp) denotes the proportion of variance reduction by feature. 
}
  \begin{tabular}{p{5cm}|r}
    \toprule
      Feature & Imp \\
    \midrule
    Relative duration \textit{(Class)--std.dev }&  0.24  \\ 
    Absolute duration \textit{(Sched.)--std.dev }&  0.19  \\   
    Relative duration \textit{(Sched.|Acad.)--std.dev} &  0.12  \\ 
    Absolute duration \textit{(Other|Acad.)--std.dev}  &  0.11  \\ 
    Relative duration \textit{(Class)--mean  } &  0.10  \\ 

\bottomrule
\end{tabular}
\label{table:feat_impo}
\end{wraptable}

Finally, to further dissect the model and understand how the collocation-based features are related to the final score, we evaluate the feature importance of the selected variables~\cite{scikit-learn}. \autoref{table:feat_impo} shows the top five features in the best model, \Mgro{} with Gradient Boost. 
It is notable that three of these capture relative behaviors (e.g., percentage of time students were present in collocation of group members). 
Another noticeable aspect is that  four of these features are based on the variance in collocations. These features essentially describe the consistency in collocation patterns (e.g., being collocated with group members every week for a fixed period of time). 
\edit{The most important feature was the relative time an individual collocated with their team during class times. Note, this differs from class attendance as project groups were expected to meet on their own during lecture times (at any location). Individuals meeting during lecture times could be considered more conscientious as it's a more responsible use of their schedule for their project. The next most important feature is the time spent in scheduled meetings. These were based on collocations during the self-reported meeting times. Since this feature captured collocation beyond lectures, it could explain the additional effort expended by teams. Lastly, we also notice that the dominant collocation behaviors are in academic spaces. This includes the collocations beyond outside lecture times and beyond scheduled meetings, i.e., ``other'' times. This could capture certain ad-hoc meetings as none of the participants expressed knowing their group members outside the class. }
\revision{Together, this indicates that coarsely described collocation behaviors of group members are associated with member performance.} 


\section{Discussion}

\revision{Our study demonstrates that coarse collocation patterns of group members, inferred from WiFi network logs, are significantly related to their performance.}
This presents new opportunities to harness archival network logs to scale analyses of social behaviors for larger groups, and potentially entire campuses. 
\revision{Arguably, approaches that employ WiFi RTLS~\cite{ciscoloc,accuwareloc} have greater spatio-temporal resolution~\cite{hong2016socialprobe,zakaria2019stressmon} for collocating group members. However, our approach is applicable to the many universities that do not install such technologies throughout their WiFi network. To access this data researchers need to collaborate with their IT departments and establish strict protocols that describe how the data will be protected (and de-identified, how often it will be provided and how long can it be accumulated).   Accessing network logs is not uncommon at universities and has no additional overheard. In fact, this alternative provides an additional benefit to universities without excessive spending or intrusion of students' privacy expectations.}
For instance, with an empirical understanding of how successful project groups 
\revision{collocate on campus, }
\revision{instructors can tailor recommendations for how project courses need to be conducted and what kind of space resources are expected from the university.}
\edit{This section illustrates other potential use-cases to harness these logs for assessing long-term social behaviors at scale.}

\subsection{Applications of Inferring Collocations for Academic Experiences}
Harnessing data already collected at the infrastructure facilitates long-term analyses of \revision{collocations} in a large cohort of students. In Section~\ref{sec:validation_results}, we show that collocation behavior of project group members \revision{is significantly related to} final scores. In other words, \revision{we present empirical results that even coarsely inferred collocation of related individuals is linked to academic outcomes.}
This enables instructors to provide data-driven insights to a new cohort based on actual behaviors of successful teams. 
\revision{However, collocation is only beneficial for certain kinds of projects~\cite{olson2000distance, kozlowski1987exploration, mark2005no}, such as software development, or, as in our case, design. To understand the transferability of our results to other forms of academic work, researchers need to further inspect what occurs between the group members during collocation. Identifying these activities can help define which characteristics of collocated synchronous interactions~\cite{olson2000distance} are actually associated with higher performance. For example, project members might just be more dedicated to their tasks in the presence of others~\cite{mark2005no}, or collocation might improve their social bond and make them more comfortable about feedback~\cite{fruchter2010tension}. }
\revision{Qualitative interviews along with momentary assessments can guide researchers to automatically infer the social importance of different collocations based on the location, time, and  history of collocated individuals.}
This knowledge could be used to augment the static semantic labels of places \revision{and instead illustrate a more dynamic social blueprint of campus.} 
\revision{Moreover, since these logs can be retroactively obtained, it can provide data to explore new questions that help determine student outcomes.}
For instance, how do members of teams with prior collocations work in comparison to teams of strangers~\cite{hasan2019prior}, or how different are \revision{collocation patterns} in a new cohort for a student from a marginalized community~\cite{prakash2017correlates}. 
\revision{Practically, these results also have implications for remote learning as more universities have embraced distributed classrooms. This helps universities consider the trade-offs for using spaces for collocated group activities while also promoting the need for remote collaboration technologies that can approximate collocation behavior --- similar to what has been advocated by the CSCW community for dispersed information work~\cite{olson2000distance,cramton2001mutual,edmondson2001disrupted,geister2006effects,hinds2003out}.}
\revision{Theoretically, our work begs to question the relationship between collocation of students and social relationships outside curricular activities. While collocation of team members can build stronger social ties~\cite{trainer2016hackathon}, it is yet to be determined if the same can be said for students not associated through projects or academic outcomes. }

\subsection{Design Implications for Other Application Domains}
\label{sec:disc-other-domains}
Since WiFi logs are archival data that can be easily collected at scale
if a mechanism to obtain broad informed consent can be devised, it can be used to answer different types of questions. \revision{Although low in resolution, it can provide insights into student behavior over entire cohorts throughout their academic tenure. Such scale has not to date been a practical opportunity for study with location data}. This paper studies the collocation of individuals known to share experiences (project teams); however, it can be used in other additional ways:

\begin{enumerate}
    \item \textit{Interaction Networks: } \edit{ Similar to attempts with other sensors~\cite{hong2016socialprobe}, collocation can be used to develop social networks representing an entire campus community.} Moreover, by contextualizing these collocations based on when and where these interactions happen, it can help study multiplexed social networks~\cite{hristova2014keep} for different purposes (e.g., residential ties for roommates, academic ties for project teams, or recreational ties for parties). 
    \item \textit{Interaction Spaces: } \edit{These logs can also help describe the utilization of spaces for social and academic purposes --- insights particularly valuable to \revision{facility} managers, urban planners, and \revision{ even epidemiologists} focusing on campus redesign based on 
    \revision{community} 
    needs and habits. Retrospective analyses 
    can describe where students congregate before exam week, or when to expect parties, to better prepare facilities and services for future semesters.}
    \item \textit{Congestion Patterns: } \edit{While this study focuses on dwelling, it also enables studies movement and pathway, e.g., between classes, and inform the design of indoor hallways and routes.}
\end{enumerate}

These possibilities also motivate the use of
collocation behaviors to inform the design of various non-academic applications for different stakeholders in domains like health and wellbeing.

\subsubsection{Mental Wellbeing}
The applications related to academic outcomes discussed earlier have implications for a student's mental wellbeing. 
\revision{Prior work provides some evidence that collocation supports psychosocial safety~\cite{edmondson1999psychological}, which enables individuals to take risks without anxiety. Similarly, collocation can also help accelerate conflict resolution~\cite{hinds2003out, cramton2001mutual}. Moreover, the presence of others  can have immediate social impact, e.g., others' progress is visible but their feedback is immediate, which makes collocation both constraining and enabling~\cite{chidambaram2005out}. However, it is yet to be investigated how collocation impacts mental wellbeing outside of work-related social ties.}
This data makes it possible to evaluate \revision{such questions}
over time, for both positive and negative outcomes. Major events on campus can impact \revision{collocation behaviors potentially linked to mental wellbeing}. This could either be a violent incident~\cite{burns1999school} (e.g., during a shooting) or an enforced lockdown (e.g., during a pandemic).
In fact, the absence of \revision{collocation could be associated with social isolation}, which in turn is related to stress, affect, and depression~\cite{skek1991life,finch1997factor,ford1990relationship}. 
Although these kinds of analyses might be hard to justify in real-time, post-hoc analysis of these trends can provide insights to support positive trends or mitigate negative ones.

\subsubsection{Physical Health}
\label{sec:disc-phy-health}
\revision{Collocation}, and the lack thereof, are important behaviors in the context of contagious diseases, something that has become very clear in 2020 with the Coronavirus Disease (COVID-19) pandemic that is affecting people globally. Literature in epidemiology provides substantial evidence that  \textit{social distancing} helps reduce the spread of influenza~\cite{glass2006targeted} and coronaviruses~\cite{martin2020effectiveness}. Even though WiFi-based collocation is too coarse to determine physical contact at the spatial resolution of 6--10 feet, 
\revision{identifying proximity in the same room}
has applications for both reactive and proactive measures. 
In terms of the former, similar processing pipelines aid contact tracing by automatically assessing the likelihood of individuals at risk based on the amount of collocation they may have had with a known contagious set of individuals. Although this can have false positives, it can still render a risk-based prioritization to help with screening during highly contagious outbreaks
Using 
\revision{collocation to construct network graphs}
campus health officials can look at historically-accumulated data to understand which students 
\revision{were in proximity of infected ones and nudge for diagnostic testing. Another potential application would be to promote cautious behaviors in students by allowing them to reflect on the degrees-of-separation between them and other positive cases in the community.}
Alternatively, similar data can be leveraged to develop and simulate proactive measures that assist campuses in resuming and continuing safe operations during a period of contagion. By modeling prior data based on congestion and pedestrian traffic, it is possible to determine the specific bottlenecks on campus that should be regulated because of risk through both direct interaction and exposure to contact surfaces (e.g., door handles at exits)~\cite{cauchemez2008estimating}. Even simpler solutions of applications that depict occupancy of spaces to students in real-time can help them adopt safer behaviors by avoiding interactions~\cite{fenichel2011adaptive}. Policies such as instituting one-way walkways, assigned seating in classrooms, hybrid physical-remote class attendance policies to reduce student density in classrooms or other creative measures can be tested \revision{on prior data} to see how much they impact the risk of exposure to an individual and an entire campus community.

\subsection{Privacy, Policy and Ethics}

The use of passive sensing technologies captured in the digital infrastructure of a campus can characterize human behavior and holds exciting potential because it can be automated and scaled. This mitigates the limitations of manual sensing such as self-reports of experiences or even requiring all individuals to install and consent to passive sensing on personal devices. However, since this paper highlights the feasibility of appropriating data archived in existing systems, it also elicits new concerns when considering practical deployments.

Any pervasive technology with the potential of large-scale passive sensing faces privacy concerns~\cite{onnela2016harnessing}. In the scope of our work, the privacy concerns can be related to both the data that is collected (coarse location) as well as what it can infer (performance) and its eventual implications ~\cite{langheinrich2009privacy}.
From the perspective of data collected, the use of the WiFi association logs is more privacy-preserving in comparison to installing an application on a client device that accumulates data to a central server. Such client-based applications can be perceived as  invasive not only because such agents can collect sensitive data---possibly more than what the user is aware of---but also because the aggregation can be continuous and unbounded, e.g., an application logs locations even beyond the campus perimeter~\cite{shilton2009four}. On the other hand, infrastructure-based localization is limited only to timestamps of network associations and does not elicit anxieties related to a client-side agent leaking data from other sensors.
Moreover, these approaches are also localized to the campus. However, automatic computation of where individuals are and whom they interact with can be considered sensitive by  students~\cite{rooksby2019student}. Therefore, when adopting such approaches to infer interactions, stakeholders need to consider approaches like \textit{differential privacy} to obfuscate sensitive data~\cite{dwork2014algorithmic}. 

Related to the privacy concerns is also establishing policy around data access. This paper and prior work showcase the importance of collocations and how it can be inferred unobtrusively. However, this involves a centralized observer that harnesses location data, and even when anonymous, this can be used to trivially identify individuals~\cite{hubaux2020decentralized}. A predator can incisively connect certain dwelling patterns if they choose to, e.g., lecture rooms can reveal a schedule and potentially an individual. To protect against this, more data can be abstracted, i.e., the AP locations can be anonymized as well (while still retaining category, floor, and relative information). Yet, it still needs to be established which people have the privileges to query for information and what the queries can be. In fact~\citet{bagdasaryan2019ancile} have proposed a system for managing the privacy of ubiquitous computing systems that limits the use of the data itself~\cite{bagdasaryan2019ancile}. Moreover, campuses can adapt existing policies regarding access to student records to protect \revision{student collocation patterns.}

Finally, we also need to discuss the ethics of such inferences. \revision{Our work studies network logs to understand collocation and performance, but it also bears implications for other applications, many of which are of operational significance, not necessarily intellectual ones (Section~\ref{sec:disc-other-domains}). Since accumulation of network association logs is not uncommon at universities, it does not present any new surveillance infrastructure and instead posits reusing existing methods.} \revision{One might also use the \textit{principle of proportionality} argument to justify that individuals collocating on campus is public information~\cite{langheinrich2009privacy}}.~\revision{However, as ~\citet{wang2009privacy} argue, it is not the localization that is sensitive, but the accumulation and aggregation of such data that makes privacy negotiations challenging~\cite{wang2009privacy}}. ~\revision{These differing expectations of how this data is used can be considered concerning by the campus community}.  ~\revision{In scenarios that induce ~\textit{shared vulnerabilities}~\cite{bruneau2020ethical}, such as the outbreak of a contagious epidemic, campus stakeholders may consider using such technology for securing public health (Section~\ref{sec:disc-phy-health}).} This can be assumed to be a low-burden resource for a campus as any individual that connects to the network effectively opts-in their data for this analysis. \revision{However, even in such scenarios of practical deployment, it is imperative for any community that seeks to use this data to secure some form of consent. For example, policy-makers can communicate a blanket opt-out consent that allows processing of anonymized data (both identity and location) to inform aggregate statistics such as space usage. And any application that involves identification, e.g., contact-tracing, must preclude explicit opt-in consent}. 

\revision{Further, while repurposing these logs is a form of data minimization, we propose that stakeholders define paradigms for ``use minimization'' --- when, what, and how much of such data can be processed for applications. In many cases, this data should only be accessed retroactively, only for public areas (which excludes housing), and span no longer than 2 weeks}. \revision{Moreover, data-owners need to be enabled with affordances that let them track and selectively opt-out certain data, just the way they can with, say, credit-card transactions~\cite{wang2009privacy}}. Yet, universities must also remain aware that for a student, choosing to not connect to the network can be considered an unfair choice that limits their right to self-determine~\cite{rossler2001wert}. Although students already can choose to connect to other networks (e.g., cellular), opting out of the campus' managed network can take away critical privileges such as access to the campus library. 
Next, on any given day, while 90\% of the students in our sample were connected to the network, the students outside coverage were invisible to key applications. This missing data can have many ramifications. Even for academic performance, if instructors can use this kind of data for intervening with certain groups during midterms, those that were left out from the opportunity of improvement. \revision{Arguably, finding ways to even the privileges for those who opt-out of such systems is beyond the scope of this work, but we do recommend establishing safeguard policies that ensure no individual is penalized for their choices~\cite{sweeney2020tracking}}. Essentially designing any application of inferring collocation at scale raises concerns of fairness and accountability, which need to be carefully considered before deploying such systems. 

\subsection{Limitations and Future Work}

The most apparent limitation of using these association logs to determine \revision{collocation} 
is its low spatio-temporal resolution. This introduces reasonable uncertainty in determining the exact location of individuals ~\cite{ware2018large,loughry2007development}. Even with a lack of precision, WiFi-based localization does have its advantages. It can be argued that such approaches (e.g., \cite{hong2016socialprobe}) provide greater insight into indoor mobility and dwelling than other scalable solutions like GPS~\cite{wang2016crowdwatch,fan2015citymomentum,kanhere2013participatory}. Yet, indoor setups present several challenges that can lead to unexpected device associations~\cite{kjaergaard2012challenges}. 
As a result, an individual could be in a room and not be associated with the physically closest AP, but rather another AP node that found a stronger signal to the client. 
This creates an opportunity to deal with this noise by modeling the probability of displaced connections. Individual dwelling and collocation could be described as a probabilistic measure based on their pathway to the location. 
Other pieces of information that could help calibrate the modeling is incorporating the size and configuration of rooms and neighborhood maps of the APs.. 
Furthermore, advanced off-the-shelf methods to study archival data can be developed to make AP nodes aware of other APs visible to a client \revision{--- similar to RTLS approaches~\cite{ciscoloc,accuwareloc}}. These additional pieces of information can still be very valuable without the need of installing applications on user phones or fingerprinting the entire campus.

\revision{For identifying collocation of group members, we assume that students devices are connected to the  network during collocation. This is a reasonable assumption since our participants comprised Computer Science majors enrolled in a design course. In contrast, this might not be true for other forms of group work. For instance, projects at hardware workshops could have extended periods where digital devices are untouched and appear disconnected. Therefore, if researchers are interested in automatically inferring collocation, it is important to be aware of the expected device use during collaboration in physical spaces. On a related note, we also assume that these portable devices are a good proxy for the presence of individuals. However, edge cases can arise, for e.g., a user fails to log out from a public device and leaves campus. Such instances add noise to the data.}

We show that collocation behavior over time 
\revision{is related to performance of group members, }
even if an instance of co-presence between two individuals does not guarantee face-to-face interactions. Theoretically, this falls in line with ideas of \textit{spatiality}~\cite{olson2000distance}---when collaborators are present near each other, they are interacting through observations and an increased sense of accountability.
\revision{However, given the retrospective nature of our work we do not know what occurs during the instances of collocation. It is possible for individuals to be collocated and yet unaware of each other's presence. And in actuality these episodes could have other unseen relationships with their performance. Future work can address this by semantically categorizing collocation to disentangle collocation periods where students are likely to be aware of each other's presence.}
\revision{Moreover}
it is yet to be explored if these notions of will translate to other social relationships. Specifically, if the principles of spatiality can be extended to identify the social groups in an unsupervised way.
\section{Conclusion}

Collocation is known to be related to individual productivity outcomes, especially in terms of information work. Even on university campuses, however, students tend to collocate, especially in the context of group projects. Therefore, one way to support academic performance is to understand the extent to which collocation and performance are related.
This paper studied the feasibility of coarse collocation leveraged from WiFi network logs to investigate this phenomenon. We established the reliability of computing collocation of students in class. Then we demonstrated how collocation behaviors of project team members are related to individual performance. Additionally, we enlisted other opportunities to apply this kind of collocation data to support the campus community. This paper motivates the use of existing infrastructure data, such as WiFi logs, to perform large-scale longitudinal analyses of collocation on campus to inform applications for academic outcomes, mental wellbeing, and physical health.

\balance{}
\bibliographystyle{ACM-Reference-Format}
\bibliography{00_main}


\begin{thebibliography}{84}


\ifx \showCODEN    \undefined \def \showCODEN     #1{\unskip}     \fi
\ifx \showDOI      \undefined \def \showDOI       #1{#1}\fi
\ifx \showISBNx    \undefined \def \showISBNx     #1{\unskip}     \fi
\ifx \showISBNxiii \undefined \def \showISBNxiii  #1{\unskip}     \fi
\ifx \showISSN     \undefined \def \showISSN      #1{\unskip}     \fi
\ifx \showLCCN     \undefined \def \showLCCN      #1{\unskip}     \fi
\ifx \shownote     \undefined \def \shownote      #1{#1}          \fi
\ifx \showarticletitle \undefined \def \showarticletitle #1{#1}   \fi
\ifx \showURL      \undefined \def \showURL       {\relax}        \fi
\providecommand\bibfield[2]{#2}
\providecommand\bibinfo[2]{#2}
\providecommand\natexlab[1]{#1}
\providecommand\showeprint[2][]{arXiv:#2}

\bibitem[\protect\citeauthoryear{Aiken and West}{Aiken and West}{1990}]%
        {aiken1990invalidity}
\bibfield{author}{\bibinfo{person}{Leona~S Aiken} {and}
  \bibinfo{person}{Stephen~G West}.} \bibinfo{year}{1990}\natexlab{}.
\newblock \showarticletitle{Invalidity of true experiments: Self-report pretest
  biases}.
\newblock \bibinfo{journal}{\emph{Evaluation review}} \bibinfo{volume}{14},
  \bibinfo{number}{4} (\bibinfo{year}{1990}), \bibinfo{pages}{374--390}.
\newblock


\bibitem[\protect\citeauthoryear{Bagdasaryan, Berlstein, Waterman, Birrell,
  Foster, Schneider, and Estrin}{Bagdasaryan et~al\mbox{.}}{2019}]%
        {bagdasaryan2019ancile}
\bibfield{author}{\bibinfo{person}{Eugene Bagdasaryan},
  \bibinfo{person}{Griffin Berlstein}, \bibinfo{person}{Jason Waterman},
  \bibinfo{person}{Eleanor Birrell}, \bibinfo{person}{Nate Foster},
  \bibinfo{person}{Fred~B Schneider}, {and} \bibinfo{person}{Deborah Estrin}.}
  \bibinfo{year}{2019}\natexlab{}.
\newblock \showarticletitle{Ancile: Enhancing Privacy for Ubiquitous Computing
  with Use-Based Privacy}. In \bibinfo{booktitle}{\emph{Proceedings of the 18th
  ACM Workshop on Privacy in the Electronic Society}}.
  \bibinfo{pages}{111--124}.
\newblock


\bibitem[\protect\citeauthoryear{Bruneau, M{\"u}ller, and Gilthorpe}{Bruneau
  et~al\mbox{.}}{2020}]%
        {bruneau2020ethical}
\bibfield{author}{\bibinfo{person}{Gabriela~Arriagada Bruneau},
  \bibinfo{person}{Vincent~C M{\"u}ller}, {and} \bibinfo{person}{Mark~S
  Gilthorpe}.} \bibinfo{year}{2020}\natexlab{}.
\newblock \showarticletitle{The ethical imperatives of the COVID 19 pandemic: a
  review from data ethics}.
\newblock  (\bibinfo{year}{2020}).
\newblock


\bibitem[\protect\citeauthoryear{Burns and Crawford}{Burns and
  Crawford}{1999}]%
        {burns1999school}
\bibfield{author}{\bibinfo{person}{Ronald Burns} {and} \bibinfo{person}{Charles
  Crawford}.} \bibinfo{year}{1999}\natexlab{}.
\newblock \showarticletitle{School shootings, the media, and public fear:
  Ingredientsfor a moral panic}.
\newblock \bibinfo{journal}{\emph{Crime, law and social change}}
  \bibinfo{volume}{32}, \bibinfo{number}{2} (\bibinfo{year}{1999}),
  \bibinfo{pages}{147--168}.
\newblock


\bibitem[\protect\citeauthoryear{Cannon-Bowers and Tannenbaum}{Cannon-Bowers
  and Tannenbaum}{[n.d.]}]%
        {cannondefining}
\bibfield{author}{\bibinfo{person}{JA Cannon-Bowers} {and} \bibinfo{person}{SI
  Tannenbaum}.} \bibinfo{year}{[n.d.]}\natexlab{}.
\newblock \showarticletitle{Defining team competencies and establishing team
  training requirements,[w:] R. Guzzo, E. Salas}.
\newblock \bibinfo{journal}{\emph{Team effectiveness and decision making in
  organizations}} (\bibinfo{year}{[n.\,d.]}), \bibinfo{pages}{333--380}.
\newblock


\bibitem[\protect\citeauthoryear{Carnevale and Probst}{Carnevale and
  Probst}{1998}]%
        {carnevale1998social}
\bibfield{author}{\bibinfo{person}{Peter~J Carnevale} {and}
  \bibinfo{person}{Tahira~M Probst}.} \bibinfo{year}{1998}\natexlab{}.
\newblock \showarticletitle{Social values and social conflict in creative
  problem solving and categorization.}
\newblock \bibinfo{journal}{\emph{Journal of personality and social
  psychology}} \bibinfo{volume}{74}, \bibinfo{number}{5}
  (\bibinfo{year}{1998}), \bibinfo{pages}{1300}.
\newblock


\bibitem[\protect\citeauthoryear{Cauchemez, Valleron, Boelle, Flahault, and
  Ferguson}{Cauchemez et~al\mbox{.}}{2008}]%
        {cauchemez2008estimating}
\bibfield{author}{\bibinfo{person}{Simon Cauchemez},
  \bibinfo{person}{Alain-Jacques Valleron}, \bibinfo{person}{Pierre-Yves
  Boelle}, \bibinfo{person}{Antoine Flahault}, {and} \bibinfo{person}{Neil~M
  Ferguson}.} \bibinfo{year}{2008}\natexlab{}.
\newblock \showarticletitle{Estimating the impact of school closure on
  influenza transmission from Sentinel data}.
\newblock \bibinfo{journal}{\emph{Nature}} \bibinfo{volume}{452},
  \bibinfo{number}{7188} (\bibinfo{year}{2008}), \bibinfo{pages}{750--754}.
\newblock


\bibitem[\protect\citeauthoryear{Chai and Draxler}{Chai and Draxler}{2014}]%
        {chai2014root}
\bibfield{author}{\bibinfo{person}{Tianfeng Chai} {and}
  \bibinfo{person}{Roland~R Draxler}.} \bibinfo{year}{2014}\natexlab{}.
\newblock \showarticletitle{Root mean square error (RMSE) or mean absolute
  error (MAE)?--Arguments against avoiding RMSE in the literature}.
\newblock \bibinfo{journal}{\emph{Geoscientific model development}}
  \bibinfo{volume}{7}, \bibinfo{number}{3} (\bibinfo{year}{2014}),
  \bibinfo{pages}{1247--1250}.
\newblock


\bibitem[\protect\citeauthoryear{Chidambaram and Tung}{Chidambaram and
  Tung}{2005}]%
        {chidambaram2005out}
\bibfield{author}{\bibinfo{person}{Laku Chidambaram} {and}
  \bibinfo{person}{Lai~Lai Tung}.} \bibinfo{year}{2005}\natexlab{}.
\newblock \showarticletitle{Is out of sight, out of mind? An empirical study of
  social loafing in technology-supported groups}.
\newblock \bibinfo{journal}{\emph{Information systems research}}
  \bibinfo{volume}{16}, \bibinfo{number}{2} (\bibinfo{year}{2005}),
  \bibinfo{pages}{149--168}.
\newblock


\bibitem[\protect\citeauthoryear{Cramton}{Cramton}{2001}]%
        {cramton2001mutual}
\bibfield{author}{\bibinfo{person}{Catherine~Durnell Cramton}.}
  \bibinfo{year}{2001}\natexlab{}.
\newblock \showarticletitle{The mutual knowledge problem and its consequences
  for dispersed collaboration}.
\newblock \bibinfo{journal}{\emph{Organization science}} \bibinfo{volume}{12},
  \bibinfo{number}{3} (\bibinfo{year}{2001}), \bibinfo{pages}{346--371}.
\newblock


\bibitem[\protect\citeauthoryear{Das, Chatterjee, Chakraborty, and Mitra}{Das
  et~al\mbox{.}}{2018}]%
        {das2018groupsense}
\bibfield{author}{\bibinfo{person}{Snigdha Das}, \bibinfo{person}{Soumyajit
  Chatterjee}, \bibinfo{person}{Sandip Chakraborty}, {and}
  \bibinfo{person}{Bivas Mitra}.} \bibinfo{year}{2018}\natexlab{}.
\newblock \showarticletitle{GroupSense: A Lightweight Framework for Group
  Identification}.
\newblock \bibinfo{journal}{\emph{IEEE Transactions on Mobile Computing}}
  \bibinfo{volume}{18}, \bibinfo{number}{12} (\bibinfo{year}{2018}),
  \bibinfo{pages}{2856--2870}.
\newblock


\bibitem[\protect\citeauthoryear{{Das Swain}, Reddy, Nies, Tay, {De Choudhury},
  and Abowd}{{Das Swain} et~al\mbox{.}}{2019}]%
        {DasSwain2019FitRoutine}
\bibfield{author}{\bibinfo{person}{Vedant {Das Swain}},
  \bibinfo{person}{Manikanta~D. Reddy}, \bibinfo{person}{Kari~Anne Nies},
  \bibinfo{person}{Louis Tay}, \bibinfo{person}{Munmun {De Choudhury}}, {and}
  \bibinfo{person}{Gregory~D. Abowd}.} \bibinfo{year}{2019}\natexlab{}.
\newblock \showarticletitle{{Birds of a Feather Clock Together: A Study of
  Person--Organization Fit Through Latent Activity Routines}}.
\newblock \bibinfo{journal}{\emph{Proc. ACM Hum.-Comput. Interact}}
  \bibinfo{number}{CSCW} (\bibinfo{year}{2019}).
\newblock


\bibitem[\protect\citeauthoryear{Dwork, Roth, et~al\mbox{.}}{Dwork
  et~al\mbox{.}}{2014}]%
        {dwork2014algorithmic}
\bibfield{author}{\bibinfo{person}{Cynthia Dwork}, \bibinfo{person}{Aaron
  Roth}, {et~al\mbox{.}}} \bibinfo{year}{2014}\natexlab{}.
\newblock \showarticletitle{The algorithmic foundations of differential
  privacy}.
\newblock \bibinfo{journal}{\emph{Foundations and Trends{\textregistered} in
  Theoretical Computer Science}} \bibinfo{volume}{9}, \bibinfo{number}{3--4}
  (\bibinfo{year}{2014}), \bibinfo{pages}{211--407}.
\newblock


\bibitem[\protect\citeauthoryear{Eagle and Pentland}{Eagle and
  Pentland}{2006}]%
        {eagle2006reality}
\bibfield{author}{\bibinfo{person}{Nathan Eagle} {and}
  \bibinfo{person}{Alex~Sandy Pentland}.} \bibinfo{year}{2006}\natexlab{}.
\newblock \showarticletitle{Reality mining: sensing complex social systems}.
\newblock \bibinfo{journal}{\emph{Personal and ubiquitous computing}}
  \bibinfo{volume}{10}, \bibinfo{number}{4} (\bibinfo{year}{2006}),
  \bibinfo{pages}{255--268}.
\newblock


\bibitem[\protect\citeauthoryear{Edmondson}{Edmondson}{1999}]%
        {edmondson1999psychological}
\bibfield{author}{\bibinfo{person}{Amy Edmondson}.}
  \bibinfo{year}{1999}\natexlab{}.
\newblock \showarticletitle{Psychological safety and learning behavior in work
  teams}.
\newblock \bibinfo{journal}{\emph{Administrative science quarterly}}
  \bibinfo{volume}{44}, \bibinfo{number}{2} (\bibinfo{year}{1999}),
  \bibinfo{pages}{350--383}.
\newblock


\bibitem[\protect\citeauthoryear{Edmondson, Bohmer, and Pisano}{Edmondson
  et~al\mbox{.}}{2001}]%
        {edmondson2001disrupted}
\bibfield{author}{\bibinfo{person}{Amy~C Edmondson}, \bibinfo{person}{Richard~M
  Bohmer}, {and} \bibinfo{person}{Gary~P Pisano}.}
  \bibinfo{year}{2001}\natexlab{}.
\newblock \showarticletitle{Disrupted routines: Team learning and new
  technology implementation in hospitals}.
\newblock \bibinfo{journal}{\emph{Administrative science quarterly}}
  \bibinfo{volume}{46}, \bibinfo{number}{4} (\bibinfo{year}{2001}),
  \bibinfo{pages}{685--716}.
\newblock


\bibitem[\protect\citeauthoryear{Eldaw, Levene, and Roussos}{Eldaw
  et~al\mbox{.}}{2018}]%
        {eldaw2018presence}
\bibfield{author}{\bibinfo{person}{Muawya Habib~Sarnoub Eldaw},
  \bibinfo{person}{Mark Levene}, {and} \bibinfo{person}{George Roussos}.}
  \bibinfo{year}{2018}\natexlab{}.
\newblock \showarticletitle{Presence analytics: making sense of human social
  presence within a learning environment}. In \bibinfo{booktitle}{\emph{2018
  IEEE/ACM 5th International Conference on Big Data Computing Applications and
  Technologies (BDCAT)}}. IEEE, \bibinfo{pages}{174--183}.
\newblock


\bibitem[\protect\citeauthoryear{Fan, Song, Shibasaki, and Adachi}{Fan
  et~al\mbox{.}}{2015}]%
        {fan2015citymomentum}
\bibfield{author}{\bibinfo{person}{Zipei Fan}, \bibinfo{person}{Xuan Song},
  \bibinfo{person}{Ryosuke Shibasaki}, {and} \bibinfo{person}{Ryutaro Adachi}.}
  \bibinfo{year}{2015}\natexlab{}.
\newblock \showarticletitle{CityMomentum: an online approach for crowd behavior
  prediction at a citywide level}. In \bibinfo{booktitle}{\emph{Proceedings of
  the 2015 ACM International Joint Conference on Pervasive and Ubiquitous
  Computing}}. \bibinfo{pages}{559--569}.
\newblock


\bibitem[\protect\citeauthoryear{Fenichel, Castillo-Chavez, Ceddia, Chowell,
  Parra, Hickling, Holloway, Horan, Morin, Perrings, et~al\mbox{.}}{Fenichel
  et~al\mbox{.}}{2011}]%
        {fenichel2011adaptive}
\bibfield{author}{\bibinfo{person}{Eli~P Fenichel}, \bibinfo{person}{Carlos
  Castillo-Chavez}, \bibinfo{person}{M~Graziano Ceddia},
  \bibinfo{person}{Gerardo Chowell}, \bibinfo{person}{Paula A~Gonzalez Parra},
  \bibinfo{person}{Graham~J Hickling}, \bibinfo{person}{Garth Holloway},
  \bibinfo{person}{Richard Horan}, \bibinfo{person}{Benjamin Morin},
  \bibinfo{person}{Charles Perrings}, {et~al\mbox{.}}}
  \bibinfo{year}{2011}\natexlab{}.
\newblock \showarticletitle{Adaptive human behavior in epidemiological models}.
\newblock \bibinfo{journal}{\emph{Proceedings of the National Academy of
  Sciences}} \bibinfo{volume}{108}, \bibinfo{number}{15}
  (\bibinfo{year}{2011}), \bibinfo{pages}{6306--6311}.
\newblock


\bibitem[\protect\citeauthoryear{Finch, Barrera~Jr, Okun, Bryant, Pool, and
  Snow-Turek}{Finch et~al\mbox{.}}{1997}]%
        {finch1997factor}
\bibfield{author}{\bibinfo{person}{John~F Finch}, \bibinfo{person}{Manuel
  Barrera~Jr}, \bibinfo{person}{Morris~A Okun}, \bibinfo{person}{William~HM
  Bryant}, \bibinfo{person}{Gregory~J Pool}, {and} \bibinfo{person}{A~Lynn
  Snow-Turek}.} \bibinfo{year}{1997}\natexlab{}.
\newblock \showarticletitle{The factor structure of received social support:
  Dimensionality and the prediction of depression and life satisfaction}.
\newblock \bibinfo{journal}{\emph{Journal of Social and Clinical Psychology}}
  \bibinfo{volume}{16}, \bibinfo{number}{3} (\bibinfo{year}{1997}),
  \bibinfo{pages}{323--342}.
\newblock


\bibitem[\protect\citeauthoryear{Ford and Procidano}{Ford and
  Procidano}{1990}]%
        {ford1990relationship}
\bibfield{author}{\bibinfo{person}{Gary~G Ford} {and} \bibinfo{person}{Mary~E
  Procidano}.} \bibinfo{year}{1990}\natexlab{}.
\newblock \showarticletitle{The relationship of self-actualization to social
  support, life stress, and adjustment}.
\newblock \bibinfo{journal}{\emph{Social Behavior and Personality: an
  international journal}} \bibinfo{volume}{18}, \bibinfo{number}{1}
  (\bibinfo{year}{1990}), \bibinfo{pages}{41--51}.
\newblock


\bibitem[\protect\citeauthoryear{Friedman}{Friedman}{2001}]%
        {friedman2001greedy}
\bibfield{author}{\bibinfo{person}{Jerome~H Friedman}.}
  \bibinfo{year}{2001}\natexlab{}.
\newblock \showarticletitle{Greedy function approximation: a gradient boosting
  machine}.
\newblock \bibinfo{journal}{\emph{Annals of statistics}}
  (\bibinfo{year}{2001}), \bibinfo{pages}{1189--1232}.
\newblock


\bibitem[\protect\citeauthoryear{Fruchter, Bosch-Sijtsema, and
  Ruohom{\"a}ki}{Fruchter et~al\mbox{.}}{2010}]%
        {fruchter2010tension}
\bibfield{author}{\bibinfo{person}{Renate Fruchter}, \bibinfo{person}{Petra
  Bosch-Sijtsema}, {and} \bibinfo{person}{Virpi Ruohom{\"a}ki}.}
  \bibinfo{year}{2010}\natexlab{}.
\newblock \showarticletitle{Tension between perceived collocation and actual
  geographic distribution in project teams}.
\newblock \bibinfo{journal}{\emph{Ai \& Society}} \bibinfo{volume}{25},
  \bibinfo{number}{2} (\bibinfo{year}{2010}), \bibinfo{pages}{183--192}.
\newblock


\bibitem[\protect\citeauthoryear{Geister, Konradt, and Hertel}{Geister
  et~al\mbox{.}}{2006}]%
        {geister2006effects}
\bibfield{author}{\bibinfo{person}{Susanne Geister}, \bibinfo{person}{Udo
  Konradt}, {and} \bibinfo{person}{Guido Hertel}.}
  \bibinfo{year}{2006}\natexlab{}.
\newblock \showarticletitle{Effects of process feedback on motivation,
  satisfaction, and performance in virtual teams}.
\newblock \bibinfo{journal}{\emph{Small group research}} \bibinfo{volume}{37},
  \bibinfo{number}{5} (\bibinfo{year}{2006}), \bibinfo{pages}{459--489}.
\newblock


\bibitem[\protect\citeauthoryear{Glass, Glass, Beyeler, and Min}{Glass
  et~al\mbox{.}}{2006}]%
        {glass2006targeted}
\bibfield{author}{\bibinfo{person}{Robert~J Glass}, \bibinfo{person}{Laura~M
  Glass}, \bibinfo{person}{Walter~E Beyeler}, {and} \bibinfo{person}{H~Jason
  Min}.} \bibinfo{year}{2006}\natexlab{}.
\newblock \showarticletitle{Targeted social distancing designs for pandemic
  influenza}.
\newblock \bibinfo{journal}{\emph{Emerging infectious diseases}}
  \bibinfo{volume}{12}, \bibinfo{number}{11} (\bibinfo{year}{2006}),
  \bibinfo{pages}{1671}.
\newblock


\bibitem[\protect\citeauthoryear{Hasan and Koning}{Hasan and Koning}{2019}]%
        {hasan2019prior}
\bibfield{author}{\bibinfo{person}{Sharique Hasan} {and}
  \bibinfo{person}{Rembrand Koning}.} \bibinfo{year}{2019}\natexlab{}.
\newblock \showarticletitle{Prior ties and the limits of peer effects on
  startup team performance}.
\newblock \bibinfo{journal}{\emph{Strategic Management Journal}}
  \bibinfo{volume}{40}, \bibinfo{number}{9} (\bibinfo{year}{2019}),
  \bibinfo{pages}{1394--1416}.
\newblock


\bibitem[\protect\citeauthoryear{Hinds and Bailey}{Hinds and Bailey}{2003}]%
        {hinds2003out}
\bibfield{author}{\bibinfo{person}{Pamela~J Hinds} {and}
  \bibinfo{person}{Diane~E Bailey}.} \bibinfo{year}{2003}\natexlab{}.
\newblock \showarticletitle{Out of sight, out of sync: Understanding conflict
  in distributed teams}.
\newblock \bibinfo{journal}{\emph{Organization science}} \bibinfo{volume}{14},
  \bibinfo{number}{6} (\bibinfo{year}{2003}), \bibinfo{pages}{615--632}.
\newblock


\bibitem[\protect\citeauthoryear{Homans}{Homans}{1974}]%
        {homans1974social}
\bibfield{author}{\bibinfo{person}{George~C Homans}.}
  \bibinfo{year}{1974}\natexlab{}.
\newblock \showarticletitle{Social behavior: Its elementary forms}.
\newblock  (\bibinfo{year}{1974}).
\newblock


\bibitem[\protect\citeauthoryear{Hong, Luo, and Chan}{Hong
  et~al\mbox{.}}{2016}]%
        {hong2016socialprobe}
\bibfield{author}{\bibinfo{person}{Hande Hong}, \bibinfo{person}{Chengwen Luo},
  {and} \bibinfo{person}{Mun~Choon Chan}.} \bibinfo{year}{2016}\natexlab{}.
\newblock \showarticletitle{Socialprobe: Understanding social interaction
  through passive wifi monitoring}. In \bibinfo{booktitle}{\emph{Proceedings of
  the 13th International Conference on Mobile and Ubiquitous Systems:
  Computing, Networking and Services}}. \bibinfo{pages}{94--103}.
\newblock


\bibitem[\protect\citeauthoryear{Hristova, Musolesi, and Mascolo}{Hristova
  et~al\mbox{.}}{2014}]%
        {hristova2014keep}
\bibfield{author}{\bibinfo{person}{Desislava Hristova}, \bibinfo{person}{Mirco
  Musolesi}, {and} \bibinfo{person}{Cecilia Mascolo}.}
  \bibinfo{year}{2014}\natexlab{}.
\newblock \showarticletitle{Keep your friends close and your facebook friends
  closer: A multiplex network approach to the analysis of offline and online
  social ties}.
\newblock \bibinfo{journal}{\emph{arXiv preprint arXiv:1403.8034}}
  (\bibinfo{year}{2014}).
\newblock


\bibitem[\protect\citeauthoryear{Hubaux}{Hubaux}{2020}]%
        {hubaux2020decentralized}
\bibfield{author}{\bibinfo{person}{Prof Hubaux}.}
  \bibinfo{year}{2020}\natexlab{}.
\newblock \emph{\bibinfo{title}{Decentralized Privacy-Preserving Proximity
  Tracing}}.
\newblock \bibinfo{thesistype}{Ph.D. Dissertation}. \bibinfo{school}{Fraunhofer
  HHI}.
\newblock


\bibitem[\protect\citeauthoryear{Hung, Englebienne, and Kools}{Hung
  et~al\mbox{.}}{2013}]%
        {hung2013classifying}
\bibfield{author}{\bibinfo{person}{Hayley Hung}, \bibinfo{person}{Gwenn
  Englebienne}, {and} \bibinfo{person}{Jeroen Kools}.}
  \bibinfo{year}{2013}\natexlab{}.
\newblock \showarticletitle{Classifying social actions with a single
  accelerometer}. In \bibinfo{booktitle}{\emph{Proceedings of the 2013 ACM
  international joint conference on Pervasive and ubiquitous computing}}.
  \bibinfo{pages}{207--210}.
\newblock


\bibitem[\protect\citeauthoryear{Jehn}{Jehn}{1997}]%
        {jehn1997qualitative}
\bibfield{author}{\bibinfo{person}{Karen~A Jehn}.}
  \bibinfo{year}{1997}\natexlab{}.
\newblock \showarticletitle{A qualitative analysis of conflict types and
  dimensions in organizational groups}.
\newblock \bibinfo{journal}{\emph{Administrative science quarterly}}
  (\bibinfo{year}{1997}), \bibinfo{pages}{530--557}.
\newblock


\bibitem[\protect\citeauthoryear{Jehn and Mannix}{Jehn and Mannix}{2001}]%
        {jehn2001dynamic}
\bibfield{author}{\bibinfo{person}{Karen~A Jehn} {and}
  \bibinfo{person}{Elizabeth~A Mannix}.} \bibinfo{year}{2001}\natexlab{}.
\newblock \showarticletitle{The dynamic nature of conflict: A longitudinal
  study of intragroup conflict and group performance}.
\newblock \bibinfo{journal}{\emph{Academy of management journal}}
  \bibinfo{volume}{44}, \bibinfo{number}{2} (\bibinfo{year}{2001}),
  \bibinfo{pages}{238--251}.
\newblock


\bibitem[\protect\citeauthoryear{Jiang, Zhang, and Tjosvold}{Jiang
  et~al\mbox{.}}{2013}]%
        {jiang2013emotion}
\bibfield{author}{\bibinfo{person}{Jane~Yan Jiang}, \bibinfo{person}{Xiao
  Zhang}, {and} \bibinfo{person}{Dean Tjosvold}.}
  \bibinfo{year}{2013}\natexlab{}.
\newblock \showarticletitle{Emotion regulation as a boundary condition of the
  relationship between team conflict and performance: A multi-level
  examination}.
\newblock \bibinfo{journal}{\emph{Journal of Organizational Behavior}}
  \bibinfo{volume}{34}, \bibinfo{number}{5} (\bibinfo{year}{2013}),
  \bibinfo{pages}{714--734}.
\newblock


\bibitem[\protect\citeauthoryear{Jiang, Zhu, Huang, and Shou}{Jiang
  et~al\mbox{.}}{2015}]%
        {jiang2015mining}
\bibfield{author}{\bibinfo{person}{Shan Jiang}, \bibinfo{person}{Xinning Zhu},
  \bibinfo{person}{Junfei Huang}, {and} \bibinfo{person}{Guochu Shou}.}
  \bibinfo{year}{2015}\natexlab{}.
\newblock \showarticletitle{Mining social groups in campus based on wireless
  detection}. In \bibinfo{booktitle}{\emph{2015 IEEE International Conference
  on Smart City/SocialCom/SustainCom (SmartCity)}}. IEEE,
  \bibinfo{pages}{285--288}.
\newblock


\bibitem[\protect\citeauthoryear{Kanhere}{Kanhere}{2013}]%
        {kanhere2013participatory}
\bibfield{author}{\bibinfo{person}{Salil~S Kanhere}.}
  \bibinfo{year}{2013}\natexlab{}.
\newblock \showarticletitle{Participatory sensing: Crowdsourcing data from
  mobile smartphones in urban spaces}. In
  \bibinfo{booktitle}{\emph{International Conference on Distributed Computing
  and Internet Technology}}. Springer, \bibinfo{pages}{19--26}.
\newblock


\bibitem[\protect\citeauthoryear{Kj{\ae}rgaard, Blunck, Godsk, Toftkj{\ae}r,
  Christensen, and Gr{\o}nb{\ae}k}{Kj{\ae}rgaard et~al\mbox{.}}{2010}]%
        {kjaergaard2010indoor}
\bibfield{author}{\bibinfo{person}{Mikkel~Baun Kj{\ae}rgaard},
  \bibinfo{person}{Henrik Blunck}, \bibinfo{person}{Torben Godsk},
  \bibinfo{person}{Thomas Toftkj{\ae}r}, \bibinfo{person}{Dan~Lund
  Christensen}, {and} \bibinfo{person}{Kaj Gr{\o}nb{\ae}k}.}
  \bibinfo{year}{2010}\natexlab{}.
\newblock \showarticletitle{Indoor positioning using GPS revisited}. In
  \bibinfo{booktitle}{\emph{International conference on pervasive computing}}.
  Springer, \bibinfo{pages}{38--56}.
\newblock


\bibitem[\protect\citeauthoryear{Kjaergaard and Nurmi}{Kjaergaard and
  Nurmi}{2012}]%
        {kjaergaard2012challenges}
\bibfield{author}{\bibinfo{person}{Mikkel~Baun Kjaergaard} {and}
  \bibinfo{person}{Petteri Nurmi}.} \bibinfo{year}{2012}\natexlab{}.
\newblock \showarticletitle{Challenges for social sensing using WiFi signals}.
  In \bibinfo{booktitle}{\emph{Proceedings of the 1st ACM workshop on Mobile
  systems for computational social science}}. \bibinfo{pages}{17--21}.
\newblock


\bibitem[\protect\citeauthoryear{Kozlowski and Hults}{Kozlowski and
  Hults}{1987}]%
        {kozlowski1987exploration}
\bibfield{author}{\bibinfo{person}{Steve~WJ Kozlowski} {and}
  \bibinfo{person}{Brian~M Hults}.} \bibinfo{year}{1987}\natexlab{}.
\newblock \showarticletitle{An exploration of climates for technical updating
  and performance}.
\newblock \bibinfo{journal}{\emph{Personnel psychology}} \bibinfo{volume}{40},
  \bibinfo{number}{3} (\bibinfo{year}{1987}), \bibinfo{pages}{539--563}.
\newblock


\bibitem[\protect\citeauthoryear{Kozlowski and Ilgen}{Kozlowski and
  Ilgen}{2006}]%
        {kozlowski2006enhancing}
\bibfield{author}{\bibinfo{person}{Steve~WJ Kozlowski} {and}
  \bibinfo{person}{Daniel~R Ilgen}.} \bibinfo{year}{2006}\natexlab{}.
\newblock \showarticletitle{Enhancing the effectiveness of work groups and
  teams}.
\newblock \bibinfo{journal}{\emph{Psychological science in the public
  interest}} \bibinfo{volume}{7}, \bibinfo{number}{3} (\bibinfo{year}{2006}),
  \bibinfo{pages}{77--124}.
\newblock


\bibitem[\protect\citeauthoryear{Kraskov, St{\"o}gbauer, and
  Grassberger}{Kraskov et~al\mbox{.}}{2004}]%
        {kraskov2004estimating}
\bibfield{author}{\bibinfo{person}{Alexander Kraskov}, \bibinfo{person}{Harald
  St{\"o}gbauer}, {and} \bibinfo{person}{Peter Grassberger}.}
  \bibinfo{year}{2004}\natexlab{}.
\newblock \showarticletitle{Estimating mutual information}.
\newblock \bibinfo{journal}{\emph{Physical review E}} \bibinfo{volume}{69},
  \bibinfo{number}{6} (\bibinfo{year}{2004}), \bibinfo{pages}{066138}.
\newblock


\bibitem[\protect\citeauthoryear{Kreyszig}{Kreyszig}{2010}]%
        {kreyszig2010advanced}
\bibfield{author}{\bibinfo{person}{Erwin Kreyszig}.}
  \bibinfo{year}{2010}\natexlab{}.
\newblock \bibinfo{booktitle}{\emph{Advanced engineering mathematics}}.
\newblock \bibinfo{publisher}{John Wiley \& Sons}.
\newblock


\bibitem[\protect\citeauthoryear{Krumpal}{Krumpal}{2013}]%
        {krumpal2013determinants}
\bibfield{author}{\bibinfo{person}{Ivar Krumpal}.}
  \bibinfo{year}{2013}\natexlab{}.
\newblock \showarticletitle{Determinants of social desirability bias in
  sensitive surveys: a literature review}.
\newblock \bibinfo{journal}{\emph{Quality \& Quantity}} \bibinfo{volume}{47},
  \bibinfo{number}{4} (\bibinfo{year}{2013}), \bibinfo{pages}{2025--2047}.
\newblock


\bibitem[\protect\citeauthoryear{Langheinrich}{Langheinrich}{2009}]%
        {langheinrich2009privacy}
\bibfield{author}{\bibinfo{person}{Marc Langheinrich}.}
  \bibinfo{year}{2009}\natexlab{}.
\newblock \showarticletitle{Privacy in ubiquitous computing}. In
  \bibinfo{booktitle}{\emph{Ubiquitous Computing}}. CRC Press Boca Raton, FL,
  \bibinfo{pages}{95--160}.
\newblock


\bibitem[\protect\citeauthoryear{Loughry, Ohland, and DeWayne~Moore}{Loughry
  et~al\mbox{.}}{2007}]%
        {loughry2007development}
\bibfield{author}{\bibinfo{person}{Misty~L Loughry}, \bibinfo{person}{Matthew~W
  Ohland}, {and} \bibinfo{person}{D DeWayne~Moore}.}
  \bibinfo{year}{2007}\natexlab{}.
\newblock \showarticletitle{Development of a theory-based assessment of team
  member effectiveness}.
\newblock \bibinfo{journal}{\emph{Educational and psychological measurement}}
  \bibinfo{volume}{67}, \bibinfo{number}{3} (\bibinfo{year}{2007}),
  \bibinfo{pages}{505--524}.
\newblock


\bibitem[\protect\citeauthoryear{Lukowicz, Pentland, and Ferscha}{Lukowicz
  et~al\mbox{.}}{2011}]%
        {lukowicz2011context}
\bibfield{author}{\bibinfo{person}{Paul Lukowicz}, \bibinfo{person}{Sandy
  Pentland}, {and} \bibinfo{person}{Alois Ferscha}.}
  \bibinfo{year}{2011}\natexlab{}.
\newblock \showarticletitle{From context awareness to socially aware
  computing}.
\newblock \bibinfo{journal}{\emph{IEEE pervasive computing}}
  \bibinfo{volume}{11}, \bibinfo{number}{1} (\bibinfo{year}{2011}),
  \bibinfo{pages}{32--41}.
\newblock


\bibitem[\protect\citeauthoryear{Mark, Gonzalez, and Harris}{Mark
  et~al\mbox{.}}{2005}]%
        {mark2005no}
\bibfield{author}{\bibinfo{person}{Gloria Mark}, \bibinfo{person}{Victor~M
  Gonzalez}, {and} \bibinfo{person}{Justin Harris}.}
  \bibinfo{year}{2005}\natexlab{}.
\newblock \showarticletitle{No task left behind? Examining the nature of
  fragmented work}. In \bibinfo{booktitle}{\emph{Proceedings of the SIGCHI
  conference on Human factors in computing systems}}.
  \bibinfo{pages}{321--330}.
\newblock


\bibitem[\protect\citeauthoryear{Martani, Lee, Robinson, Britter, and
  Ratti}{Martani et~al\mbox{.}}{2012}]%
        {martani2012enernet}
\bibfield{author}{\bibinfo{person}{Claudio Martani}, \bibinfo{person}{David
  Lee}, \bibinfo{person}{Prudence Robinson}, \bibinfo{person}{Rex Britter},
  {and} \bibinfo{person}{Carlo Ratti}.} \bibinfo{year}{2012}\natexlab{}.
\newblock \showarticletitle{ENERNET: Studying the dynamic relationship between
  building occupancy and energy consumption}.
\newblock \bibinfo{journal}{\emph{Energy and Buildings}}  \bibinfo{volume}{47}
  (\bibinfo{year}{2012}), \bibinfo{pages}{584--591}.
\newblock


\bibitem[\protect\citeauthoryear{Mart{\'\i}n-Calvo, Aleta, Pentland, Moreno,
  and Moro}{Mart{\'\i}n-Calvo et~al\mbox{.}}{2020}]%
        {martin2020effectiveness}
\bibfield{author}{\bibinfo{person}{David Mart{\'\i}n-Calvo},
  \bibinfo{person}{Alberto Aleta}, \bibinfo{person}{Alex Pentland},
  \bibinfo{person}{Yamir Moreno}, {and} \bibinfo{person}{Esteban Moro}.}
  \bibinfo{year}{2020}\natexlab{}.
\newblock \bibinfo{booktitle}{\emph{Effectiveness of social distancing
  strategies for protecting a community from a pandemic with a data driven
  contact network based on census and real-world mobility data}}.
\newblock \bibinfo{type}{{T}echnical {R}eport}. \bibinfo{institution}{Working
  paper, https://covid-19-sds. github. io (accessed April 18, 2020)}.
\newblock


\bibitem[\protect\citeauthoryear{Meng, Liu, and Striegel}{Meng
  et~al\mbox{.}}{2014}]%
        {meng2014analyzing}
\bibfield{author}{\bibinfo{person}{Lei Meng}, \bibinfo{person}{Shu Liu}, {and}
  \bibinfo{person}{Aaron Striegel}.} \bibinfo{year}{2014}\natexlab{}.
\newblock \showarticletitle{Analyzing the longitudinal impact of proximity,
  location, and personality on smartphone usage}.
\newblock \bibinfo{journal}{\emph{Computational Social Networks}}
  \bibinfo{volume}{1}, \bibinfo{number}{1} (\bibinfo{year}{2014}),
  \bibinfo{pages}{6}.
\newblock


\bibitem[\protect\citeauthoryear{Monitor}{Monitor}{2016}]%
        {accuwareloc}
\bibfield{author}{\bibinfo{person}{Accuware Wi-Fi~Location Monitor}.}
  \bibinfo{year}{2016}\natexlab{}.
\newblock
  \bibinfo{title}{\url{https://www.accuware.com/support/wi-fi-location-monitor-accuracy/}}.
\newblock
\newblock
\newblock
\shownote{Accessed: 2020-05-10.}


\bibitem[\protect\citeauthoryear{Nefzger and Drasgow}{Nefzger and
  Drasgow}{1957}]%
        {nefzger1957needless}
\bibfield{author}{\bibinfo{person}{MD Nefzger} {and} \bibinfo{person}{James
  Drasgow}.} \bibinfo{year}{1957}\natexlab{}.
\newblock \showarticletitle{The needless assumption of normality in Pearson's
  r.}
\newblock \bibinfo{journal}{\emph{American Psychologist}} \bibinfo{volume}{12},
  \bibinfo{number}{10} (\bibinfo{year}{1957}), \bibinfo{pages}{623}.
\newblock


\bibitem[\protect\citeauthoryear{Nguyen, Phung, Gupta, and Venkatesh}{Nguyen
  et~al\mbox{.}}{2013}]%
        {nguyen2013extraction}
\bibfield{author}{\bibinfo{person}{Thuong Nguyen}, \bibinfo{person}{Dinh
  Phung}, \bibinfo{person}{Sunil Gupta}, {and} \bibinfo{person}{Svetha
  Venkatesh}.} \bibinfo{year}{2013}\natexlab{}.
\newblock \showarticletitle{Extraction of latent patterns and contexts from
  social honest signals using hierarchical Dirichlet processes}. In
  \bibinfo{booktitle}{\emph{2013 IEEE International Conference on Pervasive
  Computing and Communications (PerCom)}}. IEEE, \bibinfo{pages}{47--55}.
\newblock


\bibitem[\protect\citeauthoryear{Olgu{\'\i}n, Waber, Kim, Mohan, Ara, and
  Pentland}{Olgu{\'\i}n et~al\mbox{.}}{2008}]%
        {olguin2008sensible}
\bibfield{author}{\bibinfo{person}{Daniel~Olgu{\'\i}n Olgu{\'\i}n},
  \bibinfo{person}{Benjamin~N Waber}, \bibinfo{person}{Taemie Kim},
  \bibinfo{person}{Akshay Mohan}, \bibinfo{person}{Koji Ara}, {and}
  \bibinfo{person}{Alex Pentland}.} \bibinfo{year}{2008}\natexlab{}.
\newblock \showarticletitle{Sensible organizations: Technology and methodology
  for automatically measuring organizational behavior}.
\newblock \bibinfo{journal}{\emph{IEEE Transactions on Systems, Man, and
  Cybernetics, Part B (Cybernetics)}} \bibinfo{volume}{39}, \bibinfo{number}{1}
  (\bibinfo{year}{2008}), \bibinfo{pages}{43--55}.
\newblock


\bibitem[\protect\citeauthoryear{Olson and Olson}{Olson and Olson}{2000}]%
        {olson2000distance}
\bibfield{author}{\bibinfo{person}{Gary~M Olson} {and}
  \bibinfo{person}{Judith~S Olson}.} \bibinfo{year}{2000}\natexlab{}.
\newblock \showarticletitle{Distance matters}.
\newblock \bibinfo{journal}{\emph{Human--computer interaction}}
  \bibinfo{volume}{15}, \bibinfo{number}{2-3} (\bibinfo{year}{2000}),
  \bibinfo{pages}{139--178}.
\newblock


\bibitem[\protect\citeauthoryear{Onnela and Rauch}{Onnela and Rauch}{2016}]%
        {onnela2016harnessing}
\bibfield{author}{\bibinfo{person}{Jukka-Pekka Onnela} {and}
  \bibinfo{person}{Scott~L Rauch}.} \bibinfo{year}{2016}\natexlab{}.
\newblock \showarticletitle{Harnessing smartphone-based digital phenotyping to
  enhance behavioral and mental health}.
\newblock \bibinfo{journal}{\emph{Neuropsychopharmacology}}
  \bibinfo{volume}{41}, \bibinfo{number}{7} (\bibinfo{year}{2016}),
  \bibinfo{pages}{1691--1696}.
\newblock


\bibitem[\protect\citeauthoryear{Paper}{Paper}{2014}]%
        {ciscoloc}
\bibfield{author}{\bibinfo{person}{Cisco Wi-Fi Location-Based Services 4.1
  Design Guide~White Paper}.} \bibinfo{year}{2014}\natexlab{}.
\newblock
  \bibinfo{title}{\url{https://www.cisco.com/c/en/us/td/docs/solutions/Enterprise/Mobility/WiFiLBS-DG/wifich2.html}}.
\newblock
\newblock
\newblock
\shownote{Accessed: 2020-05-10.}


\bibitem[\protect\citeauthoryear{Pedregosa, Varoquaux, Gramfort, Michel,
  Thirion, Grisel, Blondel, Prettenhofer, Weiss, Dubourg, Vanderplas, Passos,
  Cournapeau, Brucher, Perrot, and Duchesnay}{Pedregosa et~al\mbox{.}}{2011}]%
        {scikit-learn}
\bibfield{author}{\bibinfo{person}{F. Pedregosa}, \bibinfo{person}{G.
  Varoquaux}, \bibinfo{person}{A. Gramfort}, \bibinfo{person}{V. Michel},
  \bibinfo{person}{B. Thirion}, \bibinfo{person}{O. Grisel},
  \bibinfo{person}{M. Blondel}, \bibinfo{person}{P. Prettenhofer},
  \bibinfo{person}{R. Weiss}, \bibinfo{person}{V. Dubourg}, \bibinfo{person}{J.
  Vanderplas}, \bibinfo{person}{A. Passos}, \bibinfo{person}{D. Cournapeau},
  \bibinfo{person}{M. Brucher}, \bibinfo{person}{M. Perrot}, {and}
  \bibinfo{person}{E. Duchesnay}.} \bibinfo{year}{2011}\natexlab{}.
\newblock \showarticletitle{Scikit-learn: Machine Learning in {P}ython}.
\newblock \bibinfo{journal}{\emph{Journal of Machine Learning Research}}
  \bibinfo{volume}{12} (\bibinfo{year}{2011}), \bibinfo{pages}{2825--2830}.
\newblock


\bibitem[\protect\citeauthoryear{Pincus, Gladstone, and Ehrenkranz}{Pincus
  et~al\mbox{.}}{1991}]%
        {pincus1991regularity}
\bibfield{author}{\bibinfo{person}{Steven~M Pincus}, \bibinfo{person}{Igor~M
  Gladstone}, {and} \bibinfo{person}{Richard~A Ehrenkranz}.}
  \bibinfo{year}{1991}\natexlab{}.
\newblock \showarticletitle{A regularity statistic for medical data analysis}.
\newblock \bibinfo{journal}{\emph{Journal of clinical monitoring}}
  \bibinfo{volume}{7}, \bibinfo{number}{4} (\bibinfo{year}{1991}),
  \bibinfo{pages}{335--345}.
\newblock


\bibitem[\protect\citeauthoryear{Prakash, Beattie, Javalkar, Bhattacharjee,
  Ramanaik, Thalinja, Murthy, Davey, Blanchard, Watts, et~al\mbox{.}}{Prakash
  et~al\mbox{.}}{2017}]%
        {prakash2017correlates}
\bibfield{author}{\bibinfo{person}{Ravi Prakash}, \bibinfo{person}{Tara
  Beattie}, \bibinfo{person}{Prakash Javalkar}, \bibinfo{person}{Parinita
  Bhattacharjee}, \bibinfo{person}{Satyanarayana Ramanaik},
  \bibinfo{person}{Raghavendra Thalinja}, \bibinfo{person}{Srikanta Murthy},
  \bibinfo{person}{Calum Davey}, \bibinfo{person}{James Blanchard},
  \bibinfo{person}{Charlotte Watts}, {et~al\mbox{.}}}
  \bibinfo{year}{2017}\natexlab{}.
\newblock \showarticletitle{Correlates of school dropout and absenteeism among
  adolescent girls from marginalized community in north Karnataka, south
  India}.
\newblock \bibinfo{journal}{\emph{Journal of adolescence}}
  \bibinfo{volume}{61} (\bibinfo{year}{2017}), \bibinfo{pages}{64--76}.
\newblock


\bibitem[\protect\citeauthoryear{Rooksby, Morrison, and Murray-Rust}{Rooksby
  et~al\mbox{.}}{2019}]%
        {rooksby2019student}
\bibfield{author}{\bibinfo{person}{John Rooksby}, \bibinfo{person}{Alistair
  Morrison}, {and} \bibinfo{person}{Dave Murray-Rust}.}
  \bibinfo{year}{2019}\natexlab{}.
\newblock \showarticletitle{Student perspectives on digital phenotyping: The
  acceptability of using smartphone data to assess mental health}. In
  \bibinfo{booktitle}{\emph{Proceedings of the 2019 CHI Conference on Human
  Factors in Computing Systems}}. \bibinfo{pages}{1--14}.
\newblock


\bibitem[\protect\citeauthoryear{R{\"o}ssler et~al\mbox{.}}{R{\"o}ssler
  et~al\mbox{.}}{2001}]%
        {rossler2001wert}
\bibfield{author}{\bibinfo{person}{Beate R{\"o}ssler} {et~al\mbox{.}}}
  \bibinfo{year}{2001}\natexlab{}.
\newblock \bibinfo{booktitle}{\emph{Der wert des privaten}}.
\newblock \bibinfo{publisher}{Suhrkamp Frankfurt am Main}.
\newblock


\bibitem[\protect\citeauthoryear{Rummel}{Rummel}{1976}]%
        {rummel1976understanding}
\bibfield{author}{\bibinfo{person}{Rudolph~J Rummel}.}
  \bibinfo{year}{1976}\natexlab{}.
\newblock \showarticletitle{Understanding conflict and war: vol. 2: the
  conflict helix}.
\newblock \bibinfo{journal}{\emph{Bev-erly Hills: Sage}}
  (\bibinfo{year}{1976}).
\newblock


\bibitem[\protect\citeauthoryear{Safavian and Landgrebe}{Safavian and
  Landgrebe}{1991}]%
        {safavian1991survey}
\bibfield{author}{\bibinfo{person}{S~Rasoul Safavian} {and}
  \bibinfo{person}{David Landgrebe}.} \bibinfo{year}{1991}\natexlab{}.
\newblock \showarticletitle{A survey of decision tree classifier methodology}.
\newblock \bibinfo{journal}{\emph{IEEE transactions on systems, man, and
  cybernetics}} \bibinfo{volume}{21}, \bibinfo{number}{3}
  (\bibinfo{year}{1991}), \bibinfo{pages}{660--674}.
\newblock


\bibitem[\protect\citeauthoryear{Schr{\"o}der, Hoey, and Rogers}{Schr{\"o}der
  et~al\mbox{.}}{2016}]%
        {schroder2016modeling}
\bibfield{author}{\bibinfo{person}{Tobias Schr{\"o}der}, \bibinfo{person}{Jesse
  Hoey}, {and} \bibinfo{person}{Kimberly~B Rogers}.}
  \bibinfo{year}{2016}\natexlab{}.
\newblock \showarticletitle{Modeling dynamic identities and uncertainty in
  social interactions: Bayesian affect control theory}.
\newblock \bibinfo{journal}{\emph{American Sociological Review}}
  \bibinfo{volume}{81}, \bibinfo{number}{4} (\bibinfo{year}{2016}),
  \bibinfo{pages}{828--855}.
\newblock


\bibitem[\protect\citeauthoryear{Seber and Lee}{Seber and Lee}{2012}]%
        {seber2012linear}
\bibfield{author}{\bibinfo{person}{George~AF Seber} {and}
  \bibinfo{person}{Alan~J Lee}.} \bibinfo{year}{2012}\natexlab{}.
\newblock \bibinfo{booktitle}{\emph{Linear regression analysis}}.
  Vol.~\bibinfo{volume}{329}.
\newblock \bibinfo{publisher}{John Wiley \& Sons}.
\newblock


\bibitem[\protect\citeauthoryear{S{\k{e}}k}{S{\k{e}}k}{1991}]%
        {skek1991life}
\bibfield{author}{\bibinfo{person}{Helena S{\k{e}}k}.}
  \bibinfo{year}{1991}\natexlab{}.
\newblock \showarticletitle{Life stress in various domains and perceived
  effectiveness of social support.}
\newblock \bibinfo{journal}{\emph{Polish Psychological Bulletin}}
  (\bibinfo{year}{1991}).
\newblock


\bibitem[\protect\citeauthoryear{Shi, Meng, Striegel, Qiao, Koutsonikolas, and
  Challen}{Shi et~al\mbox{.}}{2016}]%
        {shi2016walk}
\bibfield{author}{\bibinfo{person}{Jinghao Shi}, \bibinfo{person}{Lei Meng},
  \bibinfo{person}{Aaron Striegel}, \bibinfo{person}{Chunming Qiao},
  \bibinfo{person}{Dimitrios Koutsonikolas}, {and} \bibinfo{person}{Geoffrey
  Challen}.} \bibinfo{year}{2016}\natexlab{}.
\newblock \showarticletitle{A walk on the client side: Monitoring enterprise
  wifi networks using smartphone channel scans}. In
  \bibinfo{booktitle}{\emph{IEEE INFOCOM 2016-The 35th Annual IEEE
  International Conference on Computer Communications}}. IEEE,
  \bibinfo{pages}{1--9}.
\newblock


\bibitem[\protect\citeauthoryear{Shilton}{Shilton}{2009}]%
        {shilton2009four}
\bibfield{author}{\bibinfo{person}{Katie Shilton}.}
  \bibinfo{year}{2009}\natexlab{}.
\newblock \showarticletitle{Four billion little brothers? Privacy, mobile
  phones, and ubiquitous data collection}.
\newblock \bibinfo{journal}{\emph{Commun. ACM}} \bibinfo{volume}{52},
  \bibinfo{number}{11} (\bibinfo{year}{2009}), \bibinfo{pages}{48--53}.
\newblock


\bibitem[\protect\citeauthoryear{Sparrowe, Liden, Wayne, and Kraimer}{Sparrowe
  et~al\mbox{.}}{2001}]%
        {sparrowe2001social}
\bibfield{author}{\bibinfo{person}{Raymond~T Sparrowe},
  \bibinfo{person}{Robert~C Liden}, \bibinfo{person}{Sandy~J Wayne}, {and}
  \bibinfo{person}{Maria~L Kraimer}.} \bibinfo{year}{2001}\natexlab{}.
\newblock \showarticletitle{Social networks and the performance of individuals
  and groups}.
\newblock \bibinfo{journal}{\emph{Academy of management journal}}
  \bibinfo{volume}{44}, \bibinfo{number}{2} (\bibinfo{year}{2001}),
  \bibinfo{pages}{316--325}.
\newblock


\bibitem[\protect\citeauthoryear{Stallings}{Stallings}{1998}]%
        {stallings1998snmp}
\bibfield{author}{\bibinfo{person}{William Stallings}.}
  \bibinfo{year}{1998}\natexlab{}.
\newblock \bibinfo{booktitle}{\emph{SNMP, SNMPv2, SNMPv3, and RMON 1 and 2}}.
\newblock \bibinfo{publisher}{Addison-Wesley Longman Publishing Co., Inc.}
\newblock


\bibitem[\protect\citeauthoryear{Sweeney}{Sweeney}{2020}]%
        {sweeney2020tracking}
\bibfield{author}{\bibinfo{person}{Yann Sweeney}.}
  \bibinfo{year}{2020}\natexlab{}.
\newblock \showarticletitle{Tracking the debate on COVID-19 surveillance
  tools}.
\newblock \bibinfo{journal}{\emph{Nature Machine Intelligence}}
  \bibinfo{volume}{2}, \bibinfo{number}{6} (\bibinfo{year}{2020}),
  \bibinfo{pages}{301--304}.
\newblock


\bibitem[\protect\citeauthoryear{Taylor and Brown}{Taylor and Brown}{1988}]%
        {taylor1988illusion}
\bibfield{author}{\bibinfo{person}{Shelley~E Taylor} {and}
  \bibinfo{person}{Jonathon~D Brown}.} \bibinfo{year}{1988}\natexlab{}.
\newblock \showarticletitle{Illusion and well-being: a social psychological
  perspective on mental health.}
\newblock \bibinfo{journal}{\emph{Psychological bulletin}}
  \bibinfo{volume}{103}, \bibinfo{number}{2} (\bibinfo{year}{1988}),
  \bibinfo{pages}{193}.
\newblock


\bibitem[\protect\citeauthoryear{Trainer, Kalyanasundaram, Chaihirunkarn, and
  Herbsleb}{Trainer et~al\mbox{.}}{2016}]%
        {trainer2016hackathon}
\bibfield{author}{\bibinfo{person}{Erik~H Trainer}, \bibinfo{person}{Arun
  Kalyanasundaram}, \bibinfo{person}{Chalalai Chaihirunkarn}, {and}
  \bibinfo{person}{James~D Herbsleb}.} \bibinfo{year}{2016}\natexlab{}.
\newblock \showarticletitle{How to hackathon: Socio-technical tradeoffs in
  brief, intensive collocation}. In \bibinfo{booktitle}{\emph{proceedings of
  the 19th ACM conference on computer-supported cooperative work \& social
  computing}}. \bibinfo{pages}{1118--1130}.
\newblock


\bibitem[\protect\citeauthoryear{Tsubouchi, Kawajiri, and Shimosaka}{Tsubouchi
  et~al\mbox{.}}{2013}]%
        {tsubouchi2013working}
\bibfield{author}{\bibinfo{person}{Kota Tsubouchi}, \bibinfo{person}{Ryoma
  Kawajiri}, {and} \bibinfo{person}{Masamichi Shimosaka}.}
  \bibinfo{year}{2013}\natexlab{}.
\newblock \showarticletitle{Working-relationship detection from fitbit sensor
  data}. In \bibinfo{booktitle}{\emph{Proceedings of the 2013 ACM conference on
  Pervasive and ubiquitous computing adjunct publication}}.
  \bibinfo{pages}{115--118}.
\newblock


\bibitem[\protect\citeauthoryear{Van~der Vegt, Emans, and Van
  De~Vliert}{Van~der Vegt et~al\mbox{.}}{2001}]%
        {van2001patterns}
\bibfield{author}{\bibinfo{person}{Gerben~S Van~der Vegt},
  \bibinfo{person}{Ben~JM Emans}, {and} \bibinfo{person}{Evert Van De~Vliert}.}
  \bibinfo{year}{2001}\natexlab{}.
\newblock \showarticletitle{Patterns of interdependence in work teams: A
  two-level investigation of the relations with job and team satisfaction}.
\newblock \bibinfo{journal}{\emph{Personnel Psychology}} \bibinfo{volume}{54},
  \bibinfo{number}{1} (\bibinfo{year}{2001}), \bibinfo{pages}{51--69}.
\newblock


\bibitem[\protect\citeauthoryear{Wang and Loui}{Wang and Loui}{2009}]%
        {wang2009privacy}
\bibfield{author}{\bibinfo{person}{Jessa~Liying Wang} {and}
  \bibinfo{person}{Michael~C Loui}.} \bibinfo{year}{2009}\natexlab{}.
\newblock \showarticletitle{Privacy and ethical issues in location-based
  tracking systems}. In \bibinfo{booktitle}{\emph{2009 IEEE International
  Symposium on Technology and Society}}. IEEE, \bibinfo{pages}{1--4}.
\newblock


\bibitem[\protect\citeauthoryear{Wang, Guo, Peng, Zhou, and Yu}{Wang
  et~al\mbox{.}}{2016}]%
        {wang2016crowdwatch}
\bibfield{author}{\bibinfo{person}{Qianru Wang}, \bibinfo{person}{Bin Guo},
  \bibinfo{person}{Ge Peng}, \bibinfo{person}{Gang Zhou}, {and}
  \bibinfo{person}{Zhiwen Yu}.} \bibinfo{year}{2016}\natexlab{}.
\newblock \showarticletitle{CrowdWatch: Pedestrian safety assistance with
  mobile crowd sensing}. In \bibinfo{booktitle}{\emph{Proceedings of the 2016
  ACM International Joint Conference on Pervasive and Ubiquitous Computing:
  Adjunct}}. \bibinfo{pages}{217--220}.
\newblock


\bibitem[\protect\citeauthoryear{Wang, Harari, Hao, Zhou, and Campbell}{Wang
  et~al\mbox{.}}{2015}]%
        {wang2015smartgpa}
\bibfield{author}{\bibinfo{person}{Rui Wang}, \bibinfo{person}{Gabriella
  Harari}, \bibinfo{person}{Peilin Hao}, \bibinfo{person}{Xia Zhou}, {and}
  \bibinfo{person}{Andrew~T Campbell}.} \bibinfo{year}{2015}\natexlab{}.
\newblock \showarticletitle{SmartGPA: how smartphones can assess and predict
  academic performance of college students}. In
  \bibinfo{booktitle}{\emph{Proceedings of the 2015 ACM international joint
  conference on pervasive and ubiquitous computing}}.
  \bibinfo{pages}{295--306}.
\newblock


\bibitem[\protect\citeauthoryear{Wang and Shao}{Wang and Shao}{2018}]%
        {wang2018understanding}
\bibfield{author}{\bibinfo{person}{Yan Wang} {and} \bibinfo{person}{Li Shao}.}
  \bibinfo{year}{2018}\natexlab{}.
\newblock \showarticletitle{Understanding occupancy and user behaviour through
  Wi-Fi-based indoor positioning}.
\newblock \bibinfo{journal}{\emph{Building Research \& Information}}
  \bibinfo{volume}{46}, \bibinfo{number}{7} (\bibinfo{year}{2018}),
  \bibinfo{pages}{725--737}.
\newblock


\bibitem[\protect\citeauthoryear{Ware, Yue, Morillo, Lu, Shang, Kamath, Bamis,
  Bi, Russell, and Wang}{Ware et~al\mbox{.}}{2018}]%
        {ware2018large}
\bibfield{author}{\bibinfo{person}{Shweta Ware}, \bibinfo{person}{Chaoqun Yue},
  \bibinfo{person}{Reynaldo Morillo}, \bibinfo{person}{Jin Lu},
  \bibinfo{person}{Chao Shang}, \bibinfo{person}{Jayesh Kamath},
  \bibinfo{person}{Athanasios Bamis}, \bibinfo{person}{Jinbo Bi},
  \bibinfo{person}{Alexander Russell}, {and} \bibinfo{person}{Bing Wang}.}
  \bibinfo{year}{2018}\natexlab{}.
\newblock \showarticletitle{Large-scale automatic depression screening using
  meta-data from wifi infrastructure}.
\newblock \bibinfo{journal}{\emph{Proceedings of the ACM on Interactive,
  Mobile, Wearable and Ubiquitous Technologies}} \bibinfo{volume}{2},
  \bibinfo{number}{4} (\bibinfo{year}{2018}), \bibinfo{pages}{1--27}.
\newblock


\bibitem[\protect\citeauthoryear{Zakaria, Balan, and Lee}{Zakaria
  et~al\mbox{.}}{2019}]%
        {zakaria2019stressmon}
\bibfield{author}{\bibinfo{person}{Camellia Zakaria}, \bibinfo{person}{Rajesh
  Balan}, {and} \bibinfo{person}{Youngki Lee}.}
  \bibinfo{year}{2019}\natexlab{}.
\newblock \showarticletitle{StressMon: Scalable Detection of Perceived Stress
  and Depression Using Passive Sensing of Changes in Work Routines and Group
  Interactions}.
\newblock \bibinfo{journal}{\emph{Proceedings of the ACM on Human-Computer
  Interaction}} \bibinfo{volume}{3}, \bibinfo{number}{CSCW}
  (\bibinfo{year}{2019}), \bibinfo{pages}{1--29}.
\newblock


\bibitem[\protect\citeauthoryear{Zou}{Zou}{2007}]%
        {zou2007toward}
\bibfield{author}{\bibinfo{person}{Guang~Yong Zou}.}
  \bibinfo{year}{2007}\natexlab{}.
\newblock \showarticletitle{Toward using confidence intervals to compare
  correlations.}
\newblock \bibinfo{journal}{\emph{Psychological methods}} \bibinfo{volume}{12},
  \bibinfo{number}{4} (\bibinfo{year}{2007}), \bibinfo{pages}{399}.
\newblock


\end{thebibliography}

\end{document}